\documentclass[aps,floats,twocolumn,pre,10pt,groupedaddress,superscriptaddress]{revtex4-1}
\usepackage{amsmath,amssymb,amsthm,amsfonts,bm,bbm,braket,enumitem,mathtools}
\usepackage[english]{babel}

\usepackage{natbib}
\usepackage{hyperref}
\hypersetup{
	pdfauthor={Cosimo Lupo and Federico Ricci-Tersenghi},
	pdftitle={Comparison of Gabay-Toulouse and de Almeida-Thouless instabilities for the spin glass XY model in a field on sparse random graphs},
	colorlinks,
	linktocpage=true,
	pdfstartpage=1,
	pdfstartview=FitV,
	breaklinks=true,
    pageanchor=true,
    pdfpagemode=UseOutlines,
    plainpages=false,
    bookmarksnumbered,
    bookmarksopen=true,
    bookmarksopenlevel=1,
    hypertexnames=true,
    pdfhighlight=/O,
    urlcolor=blue,
    linkcolor=blue,
    citecolor=blue,
}

\allowdisplaybreaks

\usepackage[utf8x]{inputenc}	
\usepackage[T1]{fontenc} 	

\usepackage{graphicx}
\usepackage{tikz,tkz-berge,tkz-graph}
\usetikzlibrary{arrows,petri,topaths,calc}
\usepackage{array,booktabs}

\DeclarePairedDelimiter{\norm}{\lVert}{\rVert}

\begin{document}

\title{Comparison of Gabay--Toulouse and de Almeida--Thouless instabilities for the spin glass XY model in a field on sparse random graphs}

\author{Cosimo Lupo}
\affiliation{Dipartimento di Fisica, Sapienza University of Rome, P.le Aldo Moro 5, I-00185 Rome, Italy}
\author{Federico Ricci-Tersenghi}
\affiliation{Dipartimento di Fisica, INFN-Sezione di Roma1, CNR-Nanotec, Sapienza University of Rome, P.le Aldo Moro 5, I-00185 Rome, Italy}
\date{\today}

\begin{abstract}
	Vector spin glasses are known to show two different kinds of phase transitions in presence of an external field: the so-called de Almeida--Thouless and Gabay--Toulouse lines. 
	While the former has been studied to some extent on several topologies (fully connected, random graphs, finite-dimensional lattices, chains with long-range interactions), the latter has been studied only in fully connected models, which however are known to show some unphysical behaviors (e.\,g. the divergence of these critical lines in the zero-temperature limit).
	Here we compute analytically both these critical lines for XY spin glasses on random regular graphs.
	We discuss the different nature of these phase transitions and the dependence of the critical behavior on the field distribution.
	We also study the crossover between the two different critical behaviors, by suitably tuning the field distribution.
\end{abstract}

\maketitle

\section{Introduction}

Vector spin glass models~\cite{BinderYoung1986, Book_MezardEtAl1987, Book_FischerHertz1991} go beyond the much more studied discrete spin glass models (e.\,g. Ising and Potts models) by taking into account also small fluctuations in spin variables. A direct consequence of this is the presence of many more soft modes even at very low temperatures, which may change the critical behavior of the model.

Compared to Ising spin glasses, analytic studies on vector spin glasses are scarce and mostly related to fully connected models~\cite{KirkpatrickSherrington1978, deAlmeidaEtAl1978, GabayToulouse1981, ToulouseGabay1981, CraggEtAl1982, CraggSherrington1982, GabayEtAl1982, ElderfieldSherrington1982a, ElderfieldSherrington1982b, NobreEtAl1989, SharmaYoung2010} (as usual finite-dimensional vector models can be studied approximately via a perturbative renormalization group at first order in $\epsilon=6-d$~\cite{HarrisEtAl1976, ChenLubensky1977, MooreBray1982}, but the outcomes from this approach are still very much debated even for the simplest Ising models~\cite{CharbonneauYaida2017}). Unfortunately, fully connected spin glass models have some undesirable features: e.\,g. the coupling strength must be scaled as $1/\sqrt{N}$ --- with $N$ being the system size --- in order to have a good thermodynamic limit, and the critical line in the temperature versus field plane diverges in the zero-temperature limit.
These unrealistic features strongly ask for the solution of the \textit{diluted} mean-field version of vector spin glass models, where coupling strength does not need to be scaled with the system size.
However, previous works on the diluted version are even scarcer~\cite{SkantzosEtAl2005, CoolenEtAl2005, BraunAspelmeier2006, MarruzzoLeuzzi2015, MarruzzoLeuzzi2016, LupoRicciTersenghi2017a}, and none of these works discusses the physics of vector spin glass models in presence of an external field.

It is worth reminding that in $m$-component vector models with $m\ge 2$ the effect of the external field may be drastically different from what happens in Ising ($m=1$) models. For example, when the external field has the same direction on each spin variable, the longitudinal and the transverse responses may be very different (and the divergence of the latter defines the Gabay--Toulouse critical line); an effect impossible to observe in spin glass models with Ising variables.

Our main aim is to understand the nature of the phase transitions taking place in presence of an external field in vector spin glass models defined on sparse random graphs (i.e.\ having a finite coordination number).
To this aim, we focus on the simplest vector spin model, namely the XY model ($m=2$), and we study the phase diagrams and the critical behavior in presence of a uniform external field and eventually of a random field extracted according to different probability distributions.

It is worth reminding that sparse random graphs do not have short loops (their density scales as $1/N$) and so chiral ordering does not play any role on these topologies. Nevertheless our results may help elucidating the importance of the chiral ordering in finite-dimensional regular lattices, since we are going to show which kind of long-range order can actually take place without the need for a nonzero chiral order parameter.

The structure of the manuscript is the following. In Section~\ref{sec:fully_conn} we summarize the main results about vector spin glasses in a field on fully connected graphs, showing the existence of two different kinds of phase transitions: the de Almeida--Thouless (dAT) one and the Gabay--Toulouse (GT) one. Then, in Section~\ref{sec:XYmodel_sparse_graphs} we define the XY model on sparse random graphs and show how to solve it via the belief-propagation algorithm. In Section~\ref{sec:GT_dAT_lines} we compute the critical lines by studying the stability of the replica symmetric solution under different types of external field, eventually recognizing them as GT or dAT critical lines. The different kinds of symmetry breaking taking place on GT and dAT critical lines are analyzed in Section~\ref{sec:symBreak}. Then, in Section~\ref{sec:intermediate} we study the crossover between GT and dAT critical behaviors. Our concluding remarks are reported in Section~\ref{sec:concl}. Finally, in Appendix~\ref{app:SK_limit} we explain with full details how to recover the replica results cited in Section~\ref{sec:fully_conn} via an alternative and simpler derivation, based on the dense limit of the belief-propagation equations, also proving the equivalence of the two approaches.

\section{The fully connected case}
\label{sec:fully_conn}

The most generic Hamiltonian of vector spin glasses in a field reads
\begin{equation}
	\mathcal{H}[\{\boldsymbol{\sigma}_i\}]=-\sum_{(i,j)}J_{ij}\,\boldsymbol{\sigma}_i\cdot\boldsymbol{\sigma}_j -\sum_i \boldsymbol{H}_i\cdot\boldsymbol{\sigma}_i
	\label{eq:H_vector_pm_J}
\end{equation}
with spins $\{\boldsymbol{\sigma}_i\}$ being $m$-dimensional vectors of unit norm. The field $\boldsymbol{H}_i$ is represented by a $m$-dimensional vector as well, while couplings $J_{ij}$'s are as usual drawn from a suitable probability distribution $\mathbb{P}_J$ with support also on negative values.

Our work focuses on the sparse topology, that turns out to provide results that are closer to the finite-dimensional case. However, we first provide a brief summary of the results already obtained in the fully connected case --- referring to Appendix~\ref{app:SK_limit} for more details ---, since they justify some choices we will make in the following.

In the scalar case ($m=1$, i.e.\ Ising spins) --- where $J_{ij}$'s are Gaussian distributed with zero mean and variance $1/N$, while field $H$ is homogeneous --- the system exhibits a paramagnetic phase for large enough values of~$H$ and~$T$, correctly described within a replica symmetric (RS) ansatz~\cite{SherringtonKirkpatrick1975}. However, such solution turns out to be unstable when crossing a well defined line in the~$H$ vs $T$ plane, named de Almeida--Thouless line~\cite{deAlmeidaThouless1978}. A distinctive feature of the dAT line $H_\text{dAT}(T)$ is the $3/2$ exponent of its expansion at small fields, $H_\text{dAT} \simeq \tau^{3/2}$ with $\tau \equiv T_c-T$. Moreover, in fully connected models, the dAT line $H_\text{dAT}(T)$ diverges in the $T\to 0$ limit (a rather unphysical feature). Below this line, the assumption of symmetry between replicas is wrong and hence a scheme of replica symmetry breaking (RSB) has to be taken into account, eventually leading to the Parisi solution~\cite{Parisi1980b}, that actually represents the correct solution, at least for models on fully connected graphs. Notice that the case of a random quenched Gaussian-distributed field does not qualitatively change the above picture~\cite{Bray1982}, since a suitable gauge transformation maps the model back to the one with strictly positive fields~\cite{Book_Nishimori2001}.

Moving to the vector case ($m \geqslant 2$), again the RS paramagnetic solution is stable for large enough~$H$ or~$T$. However, the stability of this solution now depends on the distribution of the external field, and in particular on its direction. Indeed, Gabay and Toulouse showed in Ref.~\cite{GabayToulouse1981} that the paramagnetic solution in presence of a \textit{uniform} field becomes unstable towards RSB along a critical line $H_\text{GT}(T)$ very different from the dAT line: e.\,g., at small fields it behaves as $H_\text{GT} \simeq \tau^{1/2}$. At the Gabay--Toulouse critical line, the degrees of freedom transverse to the field direction show spontaneous symmetry breaking, highlighted by a nonzero value of the transverse overlap $q_{\perp}$. The freezing of longitudinal degrees of freedom seems to occur at lower temperatures, along a line with features reminiscent of the dAT line (however, this computation would require the use of the full RSB ansatz below the GT line, not taken into account in Ref.~\cite{GabayToulouse1981}).

Later works~\cite{CraggEtAl1982, GabayEtAl1982} then showed that RSB actually involves both transverse and longitudinal degrees of freedom along the same line --- the GT one --- though in a different manner: $q_{\perp}$ suddenly shows a \textit{strong} RSB as soon as the GT line is crossed, with a strong dependence on the Parisi parameter $x$. Instead $q_{\parallel}$, i.e.\ the longitudinal overlap with respect to the direction of the field, \textit{weakly} depends on $x$ until the dAT line is crossed, when a strong RSB occurs along the field direction as well. Hence, the dAT line in vector spin glasses with a uniform field has been recognized as a crossover between a weak and a strong RSB along the longitudinal direction, rather than a sharp phase transition from a RS to a RSB region, which at variance occurs at the GT line.

The situation changes when considering a random field, where randomness can affect the field strength, its direction or both. It has been pointed out by Sharma and Young \cite{SharmaYoung2010} that the key ingredient to avoid the GT~line and hence recover the dAT line as a sharp RS-RSB phase transition also for vector spin glasses is the randomness in the direction of the external field, while the randomness in its strength is not essential. Indeed, the crucial observation is that the GT line is also linked to a breaking in the spin symmetry (the inversion symmetry with respect to the direction given by the external field), while dAT line is not linked to any change in spin symmetry. Moreover, the resulting line of RS instability turns out to have the same~$3/2$ exponent and the same features of the dAT line in the Ising case.

\section{The XY model on sparse graphs}
\label{sec:XYmodel_sparse_graphs}

Let us now move to the diluted case. Without any loss of generality, we choose to study the $m=2$ case, that is the so-called XY model \footnote{As long as $m\ge 2$, we have both longitudinal and transverse degrees of freedom, and all the possible scenarios may take place. For $m\to\infty$ the RSB phase shrinks as $1/m$, and one would need to rescale temperatures in order to get a sensible result.}. This is a particularly simple vector model, since each spin can be  described by a single \textit{continuous} degree of freedom $\theta_i\in[0,2\pi)$, that we assume to represent the direction of the vector spin $\boldsymbol{\sigma}_i$. Analogously, also the field on the $i$-th site can be described by its modulus $H_i$ and its direction $\phi_i\in[0,2\pi)$. Moreover, keeping in mind the key observation by Sharma and Young, we fix $H_i=H$ on each site and let only directions~$\{\phi_i\}$ to vary according to a suitable probability distribution $\mathbb{P}_{\phi}$. The corresponding Hamiltonian reads
\begin{equation}
	\mathcal{H}[\{\theta_i\}]=-\sum_{(ij)\in\mathcal{E}}J_{ij}\,\cos{(\theta_i-\theta_j)} -H\sum_i \cos{(\theta_i-\phi_i)}
\end{equation}
where $\mathcal{E}$ is the edge set of the interacting graph $G$. The couplings $J_{ij}$'s are random quenched variables distributed according to the symmetric bimodal distribution
\begin{equation}
	\mathbb{P}_J(J_{ij}) = \frac{1}{2}\delta(J_{ij}-J)+\frac{1}{2}\delta(J_{ij}+J)\,.
\end{equation}

Our main task is to characterize the instability of the XY model in an external field when the underlying graph is no longer a fully connected graph, but a sparse random graph~\cite{Book_Bollobas2001}. Indeed, it is well known that many results of the mean-field approach provided by fully connected topologies are not representative of what actually happens in the finite-dimensional case: among all, the lack of strong spatial heterogeneities and the impossibility of defining and studying correlation functions. Contrarily, on sparse random graphs one can naturally define distances between spins, long-range correlations and local heterogeneities.

In particular, we focus on the ensemble of Random Regular Graphs (RRG) of fixed connectivity $C=3$, namely each vertex has exactly $C=3$ neighbors. These graphs have the crucial property of being \textit{locally tree-like}, i.e.\ each neighborhood of a given site contains no loops with high probability, eventually tending to one in the thermodynamic limit. This feature allows us to invoke the Bethe approximation~\cite{Bethe1935} and hence to exploit the Belief-Propagation (BP) algorithm~\cite{Book_Pearl1988, YedidiaEtAl2003, Book_MezardMontanari2009} to solve the model.

Notice that this approach is equivalent to the RS cavity method~\cite{MezardParisi1987} and it turns out to be always correct for models defined on trees and on large enough random graphs, given the correlations between spins decay fast enough~\cite{MezardParisi2001, Book_MezardMontanari2009}. When the RS solution becomes unstable towards RSB, one can then use the ansatz based on the 1-step replica symmetry breaking (1RSB) scheme~\cite{MezardParisi2001, MezardParisi2003} (the full-RSB scheme has not been developed yet within the cavity approach~\cite{Parisi2017}).

Since our interest is in identifying critical lines between RS and RSB phases, we are going to use a RS formalism, i.e.\ the BP algorithm, focusing specifically on the stability of the BP fixed point.

In the Bethe approximation~\cite{Bethe1935}, each physical observable can be computed starting from just the one-point $\eta_i(\theta_i)$ and the two-point $\eta_{ij}(\theta_i,\theta_j)$ marginals. In turn, their computation is based on the knowledge of the cavity marginals $\{\eta_{i\to j}(\theta_i)\}$ through the following relations
\begin{subequations}
\begin{equation}
\begin{split}
	\eta_i(\theta_i) &= \frac{1}{\mathcal{Z}_i}\,e^{\,\beta H\cos{(\theta_i-\phi_i)}}\\
	&\qquad\times\prod_{k\in\partial i}\int d\theta_k\,e^{\,\beta J_{ik}\cos{(\theta_i-\theta_k)}}\,\eta_{k\to i}(\theta_k)
\end{split}
\label{eq:full}
\end{equation}
\begin{equation}
	\eta_{ij}(\theta_i,\theta_j) = \frac{1}{\mathcal{Z}_{ij}}\,e^{\,\beta J_{ij}\cos{(\theta_i-\theta_j)}}\,\eta_{i\to j}(\theta_i)\,\eta_{j\to i}(\theta_j)
\end{equation}
\end{subequations}
where $\partial i$ is the set of neighbors of the $i$-th spin, while $\mathcal{Z}_i$ and $\mathcal{Z}_{ij}$ are normalizing constants.

Cavity marginals satisfy the set of self-consistency equations going under the name of BP equations~\cite{YedidiaEtAl2003, Book_MezardMontanari2009}:
\begin{multline}
	\eta_{i\to j}(\theta_i) = \mathcal{F}[\{\eta_{k\to i}\},\{J_{ik}\},\phi_i]\equiv \frac{1}{\mathcal{Z}_{i\to j}}\,e^{\,\beta H\cos{(\theta_i-\phi_i)}}\\
  \times\prod_{k\in\partial i\setminus j}\int d\theta_k\,e^{\,\beta J_{ik}\cos{(\theta_i-\theta_k)}}\,\eta_{k\to i}(\theta_k)
	\label{eq:BP_XY_eqs}
\end{multline}
with $\mathcal{Z}_{i\to j}$ ensuring the correct normalization. The physical meaning of $\eta_{i\to j}(\theta_i)$ is that of the probability distribution of the variable $\theta_i$ in a modified graph where edge $(i,j)$ has been removed.

When there is no external field ($H=0$), the BP equations (\ref{eq:BP_XY_eqs}) are solved by the simple paramagnetic solution $\eta_{i\to j}(\theta_i)=1/(2\pi)$ for each directed edge, which turns out to be stable only above a certain critical temperature $T_c$. Slightly below $T_c$, an approximated solution can still be analytically obtained, based on a Fourier expansion~\cite{LupoRicciTersenghi2017a}. Instead, when $T \ll T_c$ or when a field is present, the BP equations~(\ref{eq:BP_XY_eqs}) need to be solved numerically.

Since we are not interested in a given realization of the quenched disorder, but rather in the average over the disorder distribution, we solve the BP equations (\ref{eq:BP_XY_eqs}) in distribution sense. In practice we look for the probability distribution of cavity marginals $P[\eta_{i\to j}]$ solving the following equation
\begin{equation}
	\begin{split}
		P[\eta_{i\to j}] &= \mathbb{E}_{G,J,\phi} \int\prod_{k=1}^{C-1}\mathcal{D}\eta_{k\to i}\,P[\eta_{k\to i}]\\
		&\qquad\times\delta\Bigl[\eta_{i\to j}-\mathcal{F}[\{\eta_{k\to i}\},\{J_{ik}\},\phi_i]\Bigr]
		\label{eq:BP_XY_eqs_distr}
	\end{split}
\end{equation}
with $\mathbb{E}_{G,J,\phi}$ indicating the average over the ensemble of RRGs with $C=3$ and over the coupling and field probability distributions. The fixed point $\{\eta^*_{i\to j}\}$ of BP self-consistency equations~(\ref{eq:BP_XY_eqs}) so becomes a fixed point for their probability distribution, $P^*[\eta]$. The advantage brought by this approach is that the set of distributional equations~(\ref{eq:BP_XY_eqs_distr}) can be efficiently solved via the Population Dynamics Algorithm (PDA), firstly introduced in Ref.~\cite{AbouChacraEtAl1973} and then revisited and refined in Refs.~\cite{MezardParisi2001, MezardParisi2003}.

A crucial issue arising when numerically solving BP equations --- both on a given instance of the quenched disorder or in the PDA approach --- regards the discretization of continuous variables. Indeed the marginals $\eta(\theta)$ are functions over the $[0,2\pi)$ interval and would in principle require an infinite number of parameters to be described. The most effective approach~\cite{LupoRicciTersenghi2017a} is to discretize such an interval in $Q$ bins of width $2\pi/Q$ each. The resulting model is no longer endowed with the $\mathrm{O}(2)$ continuous symmetry, but with the discrete $Z_Q$ symmetry, and it is known as the \textit{Q-state clock model}~\cite{NobreSherrington1986, IlkerBerker2013, IlkerBerker2014, MarruzzoLeuzzi2015, LupoRicciTersenghi2017a, CaglarBerker2017a, CaglarBerker2017b}.

In a previous work~\cite{LupoRicciTersenghi2017a} we showed that the $Q$-state clock model provides an efficient and reliable approximation of the XY model, in both the weak and the strong disorder regimes, with deviations in physical observables decreasing exponentially fast in $Q$. This result allows us to safely use $Q=64$ in numerical simulations. Notice that BP equations for the $Q$-state clock model can be numerically solved with a computational effort that scales as $O(Q^2 N)$, with $N$ being the size of the graph (or equivalently the population size $\mathcal{N}$ in the PDA approach). Hence the exponential convergence in $Q$ actually provides a huge enhancement in numerical simulations.

\section{Computing critical lines in sparse models}
\label{sec:GT_dAT_lines}

The linear stability of the fixed point $P^*[\eta]$ of~(\ref{eq:BP_XY_eqs_distr}) provides the stability of the RS ansatz. We look at the global growth rate of perturbations $\{\delta\eta_{i\to j}(\theta_i)\}$ to fixed-point cavity marginals. Such perturbations evolve according to the following equations~\cite{LupoRicciTersenghi2017a}
\begin{equation}
	\delta\eta_{i\to j} = \sum_{k\in\partial i\setminus j}\Biggl{|}\frac{\delta \mathcal{F}[\{\eta_{k\to i}\},\{J_{ik}\},\phi_i]}{\delta \eta_{k\to i}}\Biggr{|}_{\{\eta^*_{k\to i}\}}\delta\eta_{k\to i}
	\label{eq:pert}
\end{equation}
which are nothing but the linearized version of~(\ref{eq:BP_XY_eqs}). We solve these equations via PDA, evolving a population of $\mathcal{N}$ pairs $(\eta_{i\to j},\delta\eta_{i\to j})$, actually pairs of vectors of length~$Q$. We measure the global growth rate $\lambda_{\text{BP}}$ of perturbations as follows
\begin{equation}
	\lambda_{\text{BP}} \equiv \lim_{t\to\infty}\frac{1}{t\,\mathcal{N}}\sum_{(i\to j)}\ln\int |\delta\eta_{i\to j}(\theta)| d\theta 
\end{equation}
where the integral of the absolute value of the perturbation is actually performed summing over the $Q$ discrete values. So when $\lambda_{\text{BP}}$ is positive the RS fixed point is unstable, while it is stable if $\lambda_{\text{BP}}<0$. This approach is known as \textit{Susceptibility Propagation} (SuscP). Notice that, as usual in sparse models, a strong heterogeneity characterizes the population of cavity messages, with the corresponding perturbations spanning several orders of magnitude. Hence, we chose to average the logarithm of the norm of the perturbations over the population, and this in turn make the estimate of $\lambda_{\text{BP}}$ more robust and reliable.

\begin{figure}[!t]
	\centering
	\includegraphics[width=\columnwidth]{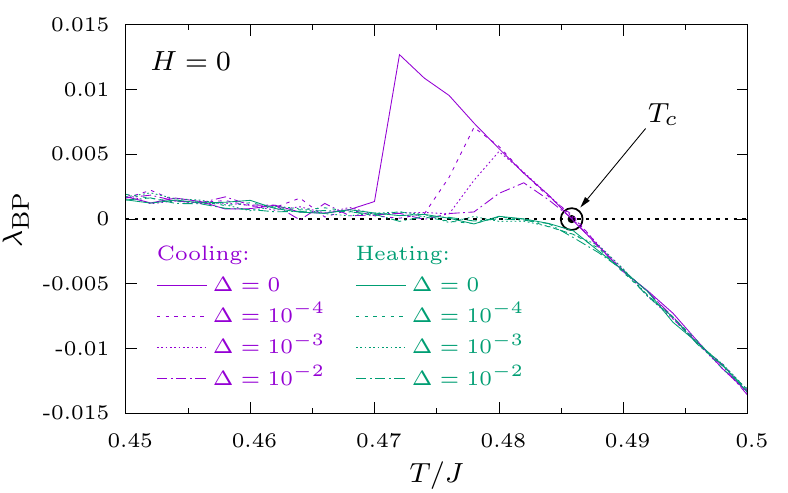}
	\caption{Stability parameter $\lambda_{\text{BP}}$ for the spin glass XY model on a $C=3$ RRG at zero field. Data are collected during cooling and heating numerical experiments with 300 iterations for temperature, and averaged over the last 150 iterations. The black dot marks the exact value for the critical temperature.}
	\label{fig:roundTrip}
\end{figure}

However, the precise determination of the critical point requires to use some precautions, because the BP equations have multiple solutions and some of these solutions (e.\,g. the paramagnetic one) change their stability at the critical point. Thus at the critical point the iterative solution of BP equations may take a large time to converge to the right solution.
In order to avoid such a critical slowing down, we solve the BP equations at a given temperature using as initial condition the fixed point reached at a nearby temperature: we call `cooling' and `heating' these two protocols to solve the BP equations, depending on whether the temperature is decreased or increased in successive rounds. Although the critical slowing down is much reduced, these two protocols have the problem that may get stuck in a solution, even when this solution becomes unstable. This is well illustrated by the cooling data at $\Delta=0$ in Fig.~\ref{fig:roundTrip}. We try to solve this problem by perturbing a little bit the initial condition before starting the iterative search for the solution to the BP equations: we add to each component of the $\eta$ marginals independent random numbers $\Delta |z|$ with $z$ being a Gaussian random variable of zero mean and unitary variance. The resulting stability parameter $\lambda_{\text{BP}}$ averaged over iterations in the time range $t\in[151,300]$ is shown in Fig.~\ref{fig:roundTrip}. We clearly see that when increasing $\Delta$, the population dynamics algorithm leaves sooner the unstable fixed point (e.\,g. the paramagnetic fixed point in the low-temperature region).

For $H=0$, a second-order phase transition occurs between the high-temperature RS-stable phase and the low-temperature RS-unstable phase, with a critical temperature $T_c=1/\beta_c$ given by \cite{SkantzosEtAl2005, CoolenEtAl2005, LupoRicciTersenghi2017a}:
\begin{equation}
	\left[\frac{I_1(\beta_c J)}{I_0(\beta_c J)}\right]^2=\frac{1}{C-1}
	\label{eq:Tc}
\end{equation}
where $C$ is the degree of the ensemble of RRG considered, while $I_0(\cdot)$ and $I_1(\cdot)$ are the modified Bessel functions of the first kind respectively of order zero and one~\cite{Book_AbramowitzStegun1964}. Critical temperatures for some values of $C$ are reported in Table~\ref{tab:Tc_vs_C}. The strength of the coupling constants $J=1/\sqrt{C-1}$ has been chosen in order to approach the critical temperature $T_c=1/2$ in the fully connected limit (indeed, when normalizing $m$-dimensional spin vectors to unity, $T_c$ is equal to $1/m$ in the fully connected limit).

\begin{table}[h]
	\caption{Critical temperatures $T_c$ for the XY model on random $C$-regular graphs with no external field and unbiased random couplings $J_{ij}\in\{+J,-J\}$. The coupling strength $J=1/\sqrt{C-1}$ has been chosen such that $\lim_{C\to\infty} T_c = 1/2$.}
	\begin{tabular}{c|c|c}
			$C$ & $T_c/J$ & $T_c$\\
		\hline
			3 & 0.4859 & 0.3436\\
			4 & 0.7012 & 0.4048\\
			6 & 0.9977 & 0.4462\\
			8 & 1.2234 & 0.4624\\
			12 & 1.5805 & 0.4765\\
			16 & 1.8704 & 0.4829\\
			20 & 2.1211 & 0.4866\\
	\end{tabular}
	\label{tab:Tc_vs_C}
\end{table}

The exact critical temperature at $H=0$ is reported in Fig.~\ref{fig:roundTrip} by a black dot.
It is clear that the best way to estimate such a critical temperature from the stability parameter $\lambda_{\text{BP}}$ is to check when the data gathered during the cooling experiment cross the axis. Such a crossing point is almost independent on the value of $\Delta$ and can be very well computed either by interpolating the data in a temperature range that includes $T_c$ or by linearly extrapolating the data collected at $T>T_c$.

On the contrary, we notice that the data in the heating experiment are of no help in identifying precisely $T_c$ for two reasons. Firstly, the stability parameter $\lambda_{\text{BP}}$ is very close to zero in a broad temperature range below~$T_c$, thus inducing a very large statistical error on the estimate of~$T_c$. Secondly, there are systematic effects that make $\lambda_{\text{BP}}$ slightly negative close to $T_c$, thus producing a biased estimate of $T_c$. A further data inspection reveals that these systematic effects are due to a very slow convergence of the population dynamics to the paramagnetic fixed point, even in presence of the $\Delta$ perturbation. In summary, a random perturbation is good for leaving the trivial fixed point, but is not as good to reach it again from a random configuration.

\begin{figure}[!t]
	\centering
	\includegraphics[width=\columnwidth]{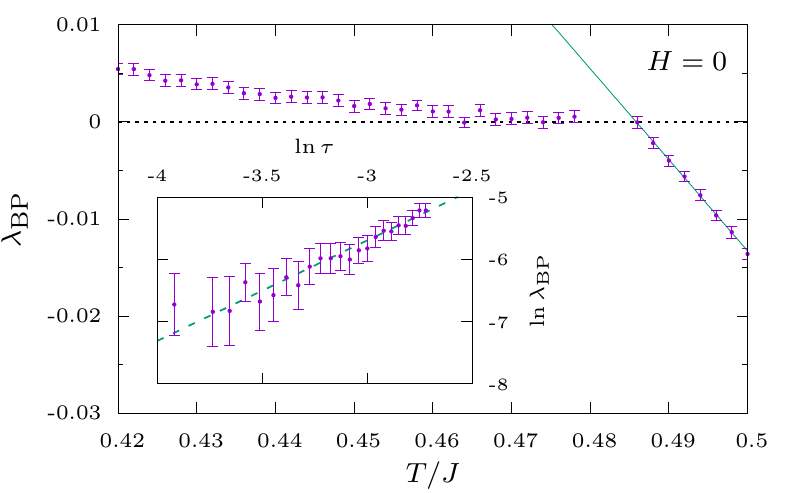}
	\caption{Stability parameter $\lambda_{\text{BP}}$ for the spin glass XY model on a $C=3$ RRG at zero field. All the points reported have been measured in the stationary regime. The full green line refers to the analytic evaluation of $\lambda_{\text{BP}}$ on the paramagnetic solution. The inset shows the power-law behavior below the critical point, $\lambda_{\text{BP}} \propto \tau^\alpha$, with $\alpha=1.6(1)$.}
	\label{fig:H0}
\end{figure}

Having discussed the possible problems arising in the numerical determination of the critical temperature, we show in Fig.~\ref{fig:H0} only the data that have been collected in the stationary regime at the stable fixed point. Some points are missing for temperatures slightly below $T_c$, but they are not really necessary in the determination of $T_c$, which is achieved by using only data with $T\ge T_c$. Being at $H=0$, we can also plot with a full line the analytic expression for $\lambda_{\text{BP}}$ that holds at the trivial paramagnetic fixed point. Instead, the behavior of the stability parameter below the critical temperature is well fitted by the power law $\lambda_{\text{BP}} \propto \tau^\alpha$ with $\alpha=1.6(1)$.

At this point, once understood how to effectively locate the transition from the RS-stable region to the RS-unstable one, we can switch on the external field. We will focus on two diametrically opposite field distributions, trying to recover also in the sparse case the well-known GT and dAT transition lines studied on fully connected graphs: firstly a uniform field and then a randomly oriented field with a flat distribution of the local field direction.

\subsection{The uniform field case}

In order to check if the GT line also appears in the sparse case, we fix the field direction to be the same on each site, e.\,g. the $\hat{x}$ direction with no loss of generality: $\mathbb{P}_{\phi}(\phi_i) = \delta(\phi_i)$.

\begin{figure}[!b]
	\centering
	\includegraphics[width=\columnwidth]{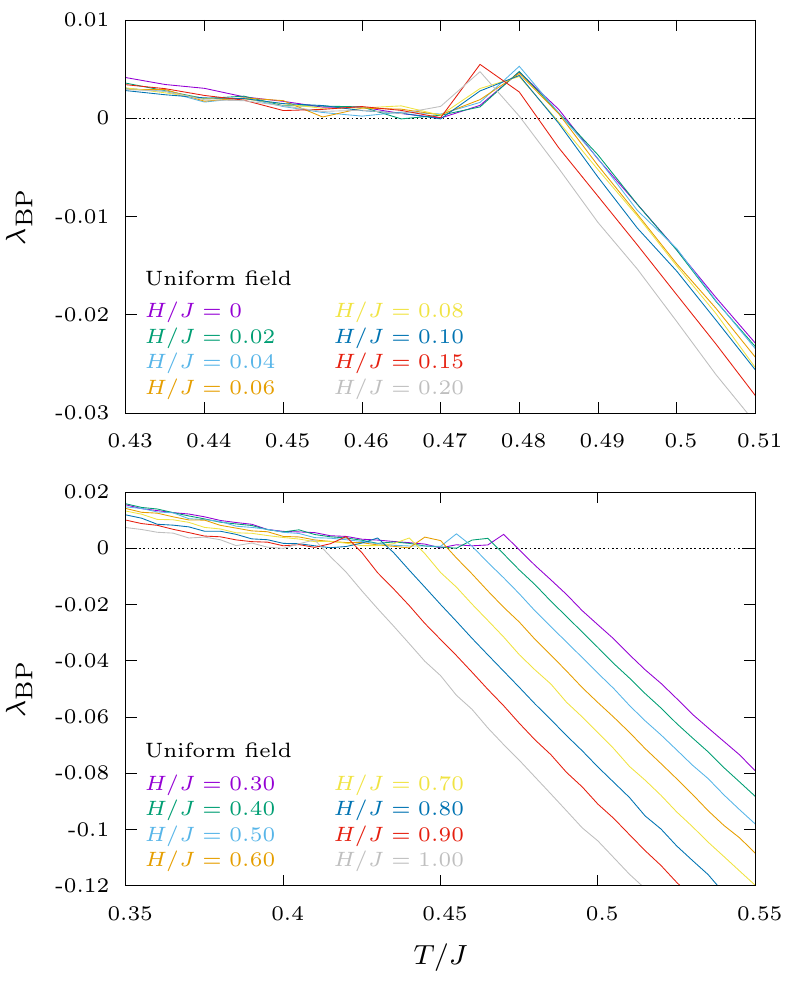}
	\caption{Stability parameter $\lambda_{\text{BP}}$ for the spin glass XY model on a $C=3$ RRG with a uniform external field of intensity $H$. The two panels show data with different ranges of fields. The lower one makes evident the leftward shift of the curves when increasing the field strength $H$.}
	\label{fig:unif_many_H}
\end{figure}

In Fig.~\ref{fig:unif_many_H} we show the stability parameter $\lambda_{\text{BP}}$ versus~$T$ with a uniform field of several intensities. We are plotting all the data collected during a cooling protocol, but from the discussion above we know that points slightly below the critical temperature should be discarded. We notice that the main effect of the field is to shift the data leftwards in the plot, that is the same instability parameter is achieved at a lower temperature.

\begin{figure}[!t]
	\centering
	\includegraphics[width=\columnwidth]{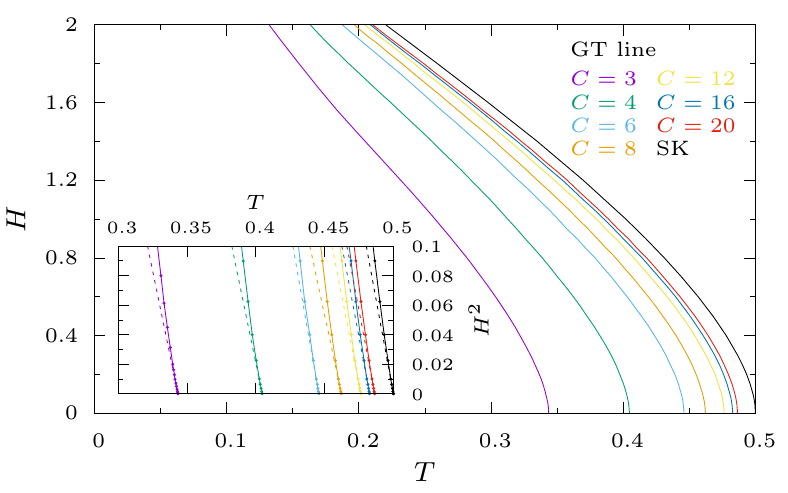}
	\caption{Critical lines in a uniform field for a spin glass XY model on a random $C$-regular graph. The corresponding line in the fully connected model (SK limit) is given by the black curve. The inset shows evidence for the $H_c(T) \propto \tau^{1/2}$ behavior, typical of the GT transition.}
	\label{fig:GT_lines}
\end{figure}

From data in Fig.~\ref{fig:unif_many_H} we estimate the critical temperature for each value of $H$ from a fit in the $T>T_c$ region. We repeat the measurements for several connectivities $C$ and we summarize in Fig.~\ref{fig:GT_lines} the results. We draw the corresponding critical lines in the $(T,H)$ plane and we observe that all they seem to have the same behavior at small fields, namely the scaling $H_c(T) \propto \tau^{1/2}$ that holds for the GT line in the fully connected model. An evidence of this is shown in the inset of Fig.~\ref{fig:GT_lines}, where we draw the critical lines in the $(T,H^2)$ plane: zooming on the interesting region of small fields, we observe a clear linear behavior in $\tau$ (such a linear behavior is soon lost due to the fact the $H_c(T)$ curves change concavity at moderately small field values). Notice that no error bars have been reported in the main plot of Fig.~\ref{fig:GT_lines}, because they would have not been appreciable, since critical points have been estimated with a statistical error of order $O(10^{-4})$.

Together with the critical curves for the diluted case with different connectivities, we also report the GT line for the fully connected graph (i.e.\ in the SK limit), computed as explained in Appendix~\ref{app:SK_limit}. It is evident the collapse of the former ones onto the latter one in the large-$C$ limit, with the most important dependence in $C$ being in the location of the zero-field critical point, while the functional form of the instability line seems to have already converged to the dense limit. So we can safely identify the critical lines reported for different $C$ values as the corresponding GT transition lines.

\subsection{The random field case}

In order to study the onset of the dAT instability in the disordered XY model, and following the suggestion of Ref.~\cite{SharmaYoung2010}, we now consider the model where the external field is constant in intensity, but has random directions~$\{\phi_i\}$ uniformly drawn in $[0,2\pi)$.

Since the field has a different (random) direction on each site, it is no longer possible to define global order parameters respectively parallel and perpendicular to the field direction; in other words, the overlaps $q_{\parallel}$ and $q_{\perp}$, used in the replica calculation to define the GT instability (see Appendix~\ref{app:SK_limit}), are now useless.
Eventually it will be possible to define the instabilities parallel and perpendicular to the field direction only locally, as it will be discussed in the next section.
For the moment, we study the global growth rate of perturbations to the BP fixed point, averaged over the population, that is the SuscP algorithm.

\begin{figure}[!t]
	\centering
	\includegraphics[width=\columnwidth]{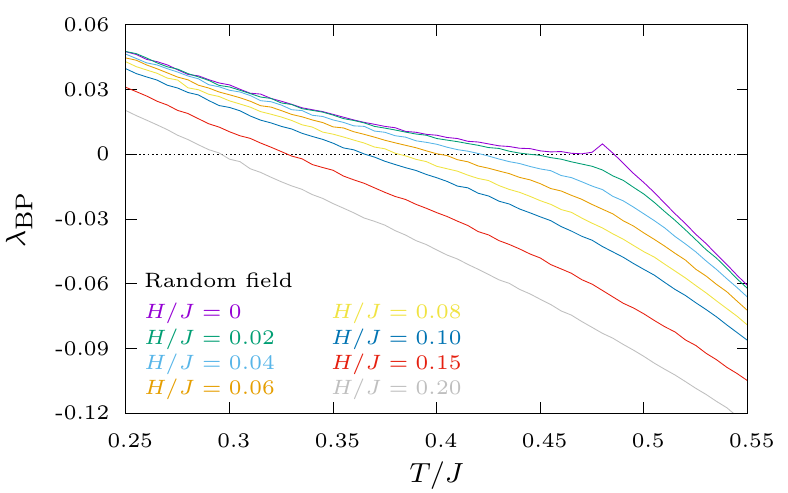}
	\caption{Stability parameter $\lambda_{\text{BP}}$ for the spin glass XY model on a $C=3$ RRG with a randomly oriented external field of fixed intensity $H$. At variance with the uniform-field case, the curve $\lambda_{\text{BP}}(T)$ mainly moves downwards when increasing~$H$, while smoothing away the zero-field singularity.}
	\label{fig:dAT_many_H}
\end{figure}

\begin{figure}[!b]
	\centering
	\includegraphics[width=\columnwidth]{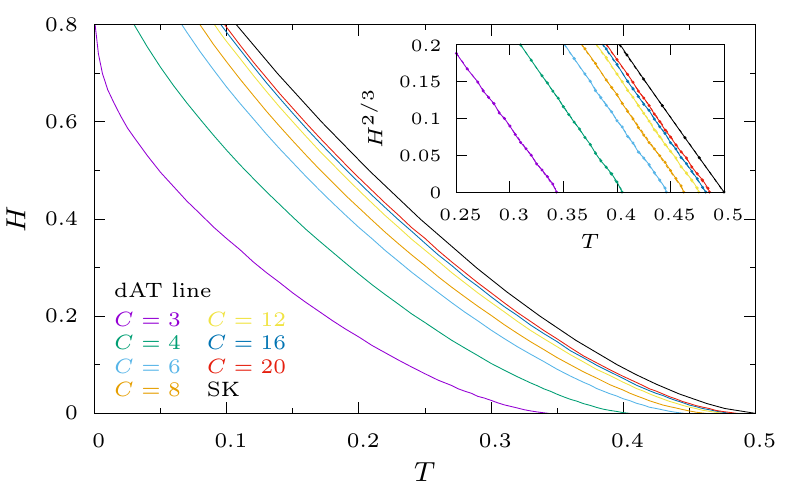}
	\caption{Critical lines in a field of random direction for a spin glass XY model a random $C$-regular graph. The corresponding line in the fully connected model (SK limit) is given by the black curve. The inset shows evidence for the $H_c(T) \propto \tau^{3/2}$ behavior, typical of the dAT transition.}
	\label{fig:dAT_lines}
\end{figure}

In Fig.~\ref{fig:dAT_many_H} we show the instability parameter $\lambda_{\text{BP}}$ versus the temperature for several values of the field intensity~$H$. At variance with the uniform-field case, now the curve moves mostly downwards with $H$ in the entire low-temperature region. The most dramatic effect, with respect to the uniform-field case, is that the stability parameter $\lambda_{\text{BP}}$ changes a lot even for very small fields, smoothing away the zero-field singularity (compare Fig.~\ref{fig:dAT_many_H} with the lower panel in Fig.~\ref{fig:unif_many_H}).

In Fig.~\ref{fig:dAT_lines} we plot the corresponding critical lines in the $(T,H)$ plane for different connectivities $C$. Close to the respective zero-field critical points, the behavior is clearly $H_c(T) \propto \tau^{3/2}$, typical of the dAT line. Again, a fast convergence towards the SK limit (the black line, computed via equations in Appendix~\ref{app:SK_limit}) can be detected, with the most important dependence in $C$ given by the location of $T_c(H=0)$. The evidence for the dAT-like behavior of these $H_c(T)$ lines is shown in the inset of Fig.~\ref{fig:dAT_lines}, where critical lines are plotted in the $(T,H^{2/3})$ plane, following the expected linear trend.

\section{GT vs dAT: different ways of breaking the spin symmetries}
\label{sec:symBreak}

\begin{figure}[!b]
	\centering
	\includegraphics[width=\columnwidth]{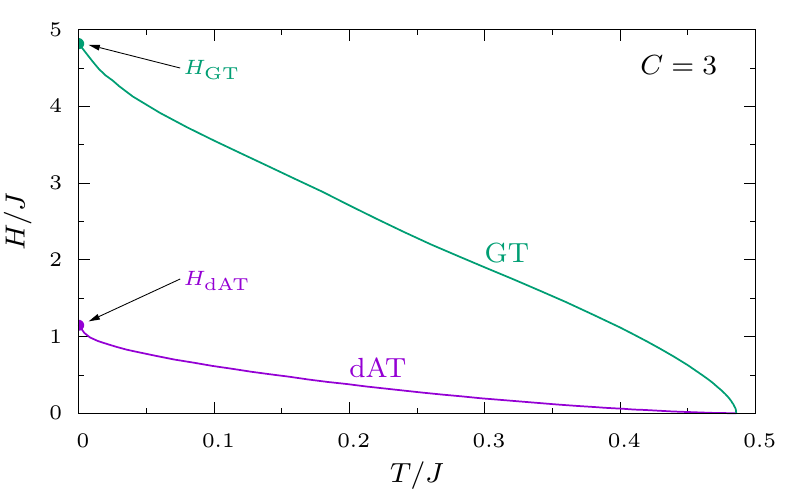}
	\caption{GT and dAT critical lines computed in the XY model with $J_{ij}=\pm1$ on a $C=3$ random regular graph.}
	\label{fig:GT_vs_dAT}
\end{figure}

In Fig.~\ref{fig:GT_vs_dAT} we show together the GT and the dAT critical lines for the XY model on a $C=3$ RRG. As explained in the previous Section, the GT line has been computed by applying a uniform field with constant direction, while the dAT line has been obtained applying a uniform field of random directions.  The overall shape of the two critical lines, including the exponent relating $H$ to $\tau$ in the vicinity of the zero-field critical point, is very similar to the fully connected case. The main difference with respect to the fully connected case is the lack of a divergence of the critical fields in the $T\to 0$ limit, as expected for the diluted case. An estimate of them, say~$H_{\text{GT}}$ and~$H_{\text{dAT}}$ respectively, can be obtained via an extrapolation from the finite-temperature datasets, though quite noisy due to the diverging slope of the two critical curves close to the $T=0$ axis. A more precise and reliable location of~$H_{\text{GT}}$ and~$H_{\text{dAT}}$ can be achieved directly in the zero-temperature setting~\cite{LupoRicciTersenghi2017a}; however, the zero-temperature BP approach requires some further precautions about the way perturbations are iteratively computed, both in the PDA~\cite{LupoRicciTersenghi2017a} as well as on a given instance of the model~\cite{[][{; C. Lupo, G. Parisi, and F. Ricci-Tersenghi (work in progress).}]Thesis_Lupo2017}.

We are now interested in understanding which symmetries get broken along these two different critical lines.
In fully connected models, the relation between the GT transition line and the freezing of the transverse degrees of freedom of spins with respect to the direction of the field is already known since the original work of Gabay and Toulouse~\cite{GabayToulouse1981}. Indeed, it is a transition from the solution $q_{\perp}=0$ to the one $q_{\perp}\neq 0$. At the same time, the dAT line --- later interpreted as a crossover --- has been naturally linked to the freezing of the longitudinal degrees of freedom. However, the strong connection between these instabilities and the distribution of the \textit{direction} of the field has been pointed out only recently by Sharma and Young \cite{SharmaYoung2010}.

Here we want to reach a deeper understanding of the kind of instabilities becoming critical on the GT and dAT lines.
To this aim, we perform a local analysis by computing, for each spin, the direction along which the most probable fluctuation may take place. We are interested in understanding whether this local fluctuation is parallel or perpendicular to the external field on the same spin (remind that in the random case the field direction changes from spin to spin and so the projection according to any global direction would be useless).

\begin{figure*}[!t]
	\centering
	\includegraphics[width=\textwidth]{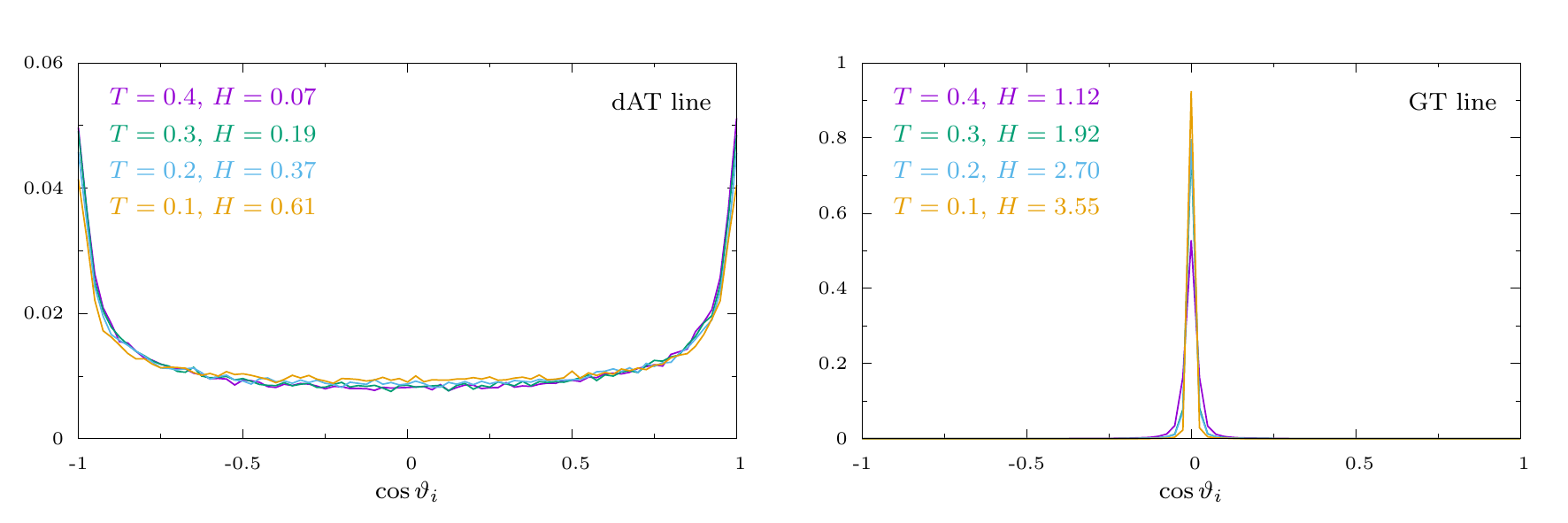}
	\caption{Probability distribution of $\cos{\vartheta_i}$ over the BP fixed-point population $\mathbb{P}^*[\eta]$ for several points along the dAT line (left panel) and the GT line (right panel). For the definition of $\vartheta_i$ see Eq.~(\ref{eq:def_cosVarTheta}) and the main text. Here $C=3$ and $J=1$.}
	\label{fig:histo_1d_ScalProd}
\end{figure*}

In the PDA we store $\mathcal{N}$ pairs $(\eta_{i\to j},\delta\eta_{i\to j})$ of cavity marginals and corresponding (linear) perturbations. Once the BP fixed point $\mathbb{P}^*[\eta]$ for the cavity marginals has been reached, the perturbations provide the direction along which such fixed point gets most easily destabilized.
Then our analysis proceeds spin by spin. For each spin $i$, we extract randomly $C$ pairs from the fixed-point population, we compute the full marginal $\eta_i$ by using Eq.~(\ref{eq:full}) and the corresponding perturbation $\delta\eta_i$ by using Eq.~(\ref{eq:pert}) with the sum running over the same $C$ randomly chosen elements.
The following local vectors
\begin{subequations}
\begin{equation}
	\boldsymbol{m}_i \equiv \int \, d\theta_i \, \eta_i(\theta_i) \, \Bigl(\cos{\theta_i},\sin{\theta_i}\Bigr)
\end{equation}
\begin{equation}
	\delta\boldsymbol{m}_i \equiv \int \, d\theta_i \, \delta\eta_i(\theta_i) \, \Bigl(\cos{\theta_i},\sin{\theta_i}\Bigr)
\end{equation}
\end{subequations}
provide the required information: $\boldsymbol{m}_i$ is the local magnetization, while $\delta\boldsymbol{m}_i$ points along the direction of the most probable local fluctuation.
The scalar product between $\delta\boldsymbol{m}_i$ and the field $\boldsymbol{H}_i$ on the same spin makes explicit the kind of perturbation to the BP fixed point: indeed a transverse perturbation would yield a scalar product close to zero, while a longitudinal perturbation would correspond to a scalar product close to one (in absolute value).
In order to be more quantitative, let us define the following local parameter
\begin{equation}
	\cos{\vartheta_i} \equiv \frac{\delta\boldsymbol{m}_i\cdot\boldsymbol{H}_i}{\norm{\delta\boldsymbol{m}_i}\norm{\boldsymbol{H}_i}} = \frac{\delta\boldsymbol{m}_i\cdot\boldsymbol{H}_i}{\delta m_i\,H}
	\label{eq:def_cosVarTheta}
\end{equation}
and let us compute its distribution by using the SuscP algorithm. Its distribution for several points along the dAT and the GT lines is depicted in Fig.~\ref{fig:histo_1d_ScalProd} for a $C=3$ RRG.

\begin{figure*}[!t]
	\centering
	\includegraphics[width=\textwidth]{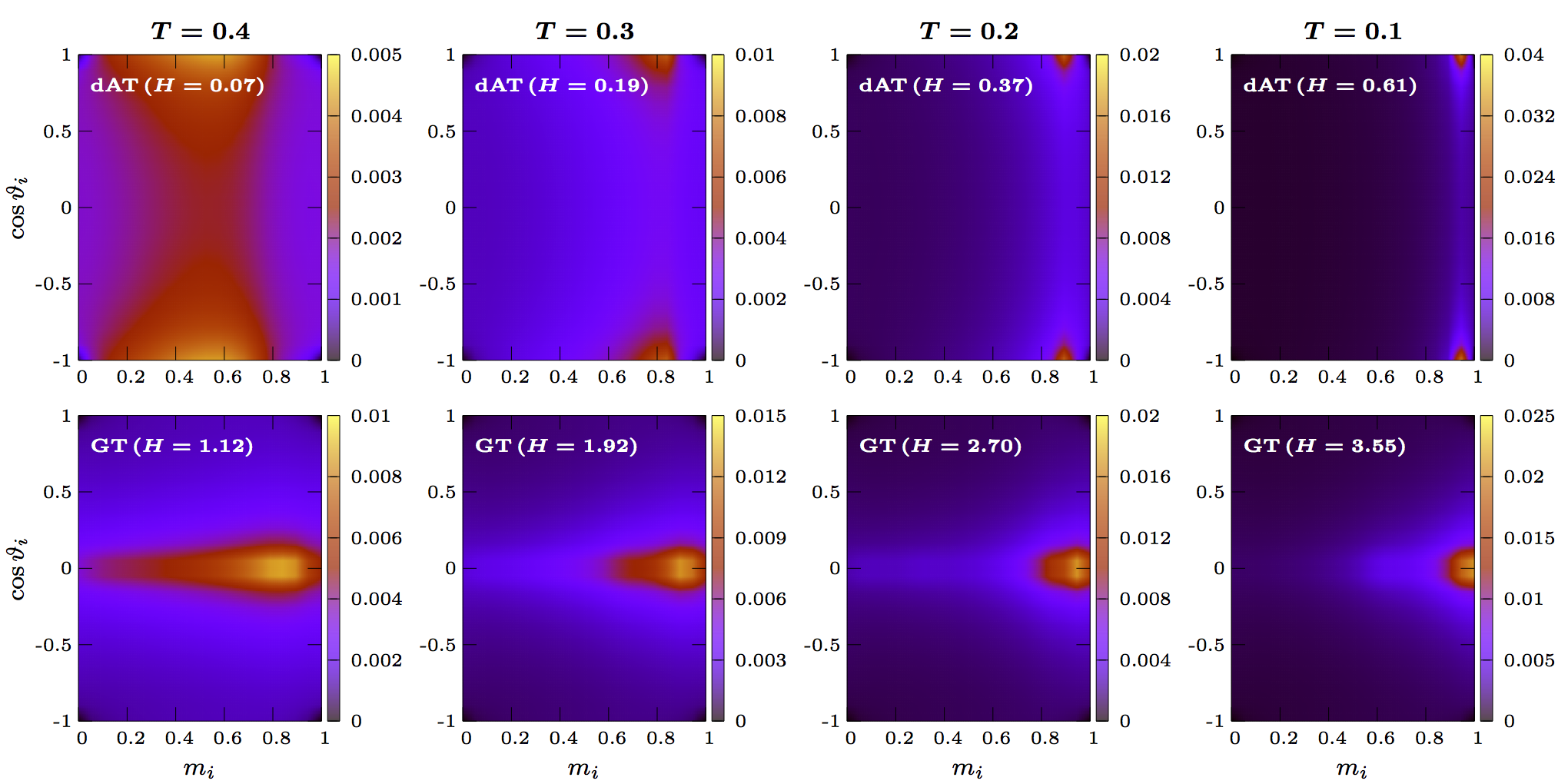}
	\caption{Joint probability distribution of $m_i$ and $\cos{\vartheta_i}$ for the same points of Fig.~\ref{fig:histo_1d_ScalProd}, better highlighting the different spin symmetries broken on dAT and GT lines, respectively. Here $C=3$ and $J=1$.}
	\label{fig:histo_2d_ScalProd}
\end{figure*}

The interpretation of the GT line as an instability in the transverse direction and that of the dAT line as an instability in the longitudinal direction --- with respect to the direction of the local field $\boldsymbol{H}_i$ --- is quite well confirmed by the two histograms of $\cos{\vartheta_i}$. Notice that the occurrence of transverse excitations also on the dAT line --- even though with a smaller probability with respect to longitudinal excitations --- is due to the fact that the field strength $H$ is not so large along such line, hence the energy cost of a transverse perturbation is surely larger than the cost of a longitudinal perturbation, but not enough to suppress them. On the other hand, on the GT line the higher the field strength $H$, the stronger the transverse behavior of perturbations.

\begin{figure*}[!t]
	\centering
	\includegraphics[width=\textwidth]{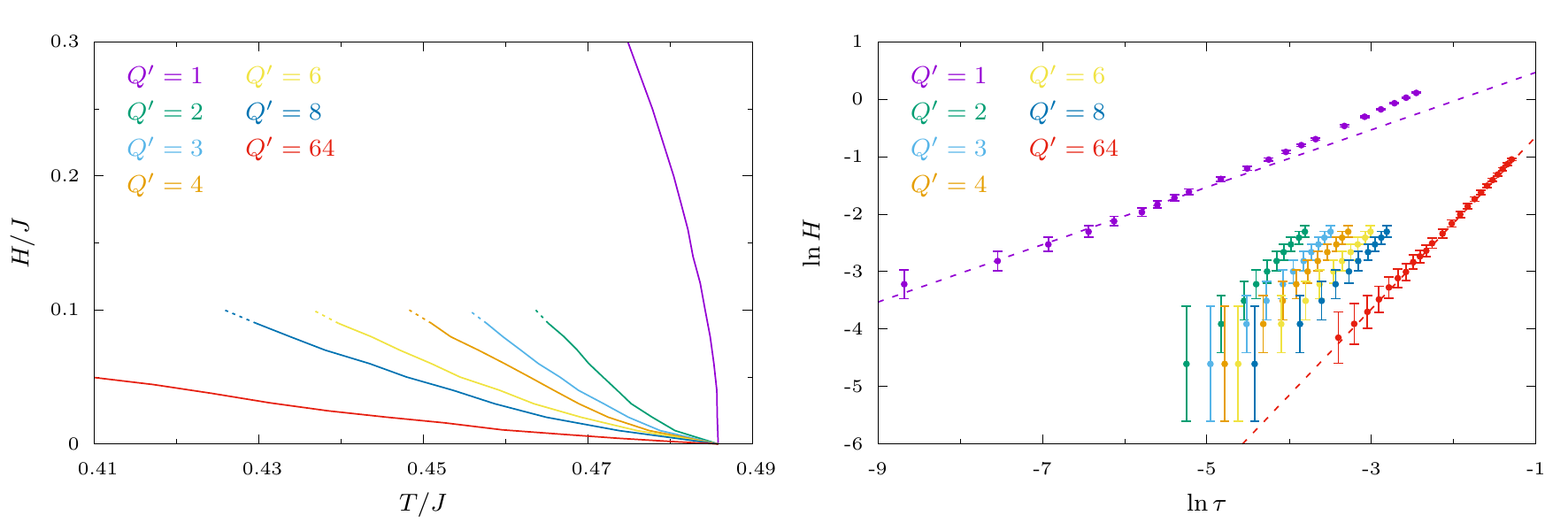}
	\caption{Critical lines in the $(T,H)$ plane for the spin glass XY model on a $C=3$ RRG with field directions $\phi=2\pi\kappa/Q$ with $\kappa\in\{0,1,\dots,Q'-1\}$ uniformly. The choice $Q'=1$ corresponds to the GT line, while $Q'=Q=64$ gives back the dAT line. Data in the right panel seem to suggest a dAT-like critical behavior for any $Q'>1$ (dashed lines have slopes $1/2$ and $3/2$, respectively).}
	\label{fig:firstClass}
\end{figure*}

The two different behaviors can be better appreciated if discriminated according to the strength of the local \textit{effective} field, given by the sum of the local field $\boldsymbol{H}_i$ and of the messages coming from the nearest-neighbor spins. A simple estimate of this strength is given by the polarization of the site marginal, namely by the modulus of the site magnetization $\boldsymbol{m}_i$. Indeed, a value of $m_i$ close to zero is representative of a weak local effective field, hence of a spin that can be easily excited along different directions with almost the same energetic cost. Instead, a strongly polarized spin is identified by a local magnetization $m_i$ close to one, hence the most likely perturbation is of course the most energetically favorable one.

In Fig.~\ref{fig:histo_2d_ScalProd} we report the joint probability distribution of $(m_i,\cos{\vartheta}_i)$ for the same points of Fig.~\ref{fig:histo_1d_ScalProd} along both instability lines. Again the difference between the basic behaviors of GT and dAT lines is quite clear, with a preference for $\cos{\vartheta_i}=0$ in the former case and for $\cos{\vartheta_i}=\pm 1$ in the latter case. In addition to this, also the dependence on the specific point of the line is evident. Indeed, when temperature is large, the local effective field is typically weak and hence the energetic cost of the two kinds of excitations is similar. So on the GT line we can also observe a nonnegligible fraction of longitudinal perturbations, conversely on the dAT line. Instead, when lowering the temperature and hence getting closer to the $T=0$ axis, the site marginals strongly polarize ($m_i\to 1$) and hence likely perturbations become more and more energetically favorable with respect to the unlikely ones. This results in well defined peaks for both lines, with the probability of an unlikely perturbation going to zero with~$T$.

So the correspondence between the two transitions in field and the breaking of spin symmetries is well established, as well as the simultaneous breaking of replica symmetry in both cases.

\section{Intermediate behaviors}
\label{sec:intermediate}

The two cases analyzed so far --- a constant field for the GT line and a random field with a flat distribution of the field local direction for the dAT line --- represent the two extremal cases in the distribution of the field direction (always keeping in mind that the field strength can be safely set equal to $H$ for all the sites without any loss of generality). Now we want to discuss some intermediate cases, in order to check which instability, between the GT-like and the dAT-like, is the dominant one in a more general case.

Since we actually solve the $Q$-state clock model, we prefer to work with probability distributions of the field direction $\phi$ taking values in the discrete set of $Q$ elements $\mathcal{S}=\{0,2\pi/Q,\ldots,2\pi(Q-1)/Q\}$.
There are still infinitely many distributions that interpolate between a delta function in $\phi=0$ and a uniform distribution over~$\mathcal{S}$.
For convenience, let us make a change of variables, taking $\phi = 2\pi\kappa/Q$ with $\kappa$ being an integer number in the range $0\le\kappa<Q$.
We choose to work with the following two classes of distributions parametrized by a single number:
\begin{itemize}
	\item $0 \le \kappa < Q'$ uniformly with probability $1/Q'$;
	\item $\kappa = 0$ with probability $1-w(Q-1)/Q$ and\\ $0<\kappa<Q$ uniformly with probability $w/Q$.
\end{itemize}
The ranges for the two parameters are $1\le Q' \le Q$ in the first class, with $Q'$ integer, and $0 \le w \le 1$ in the second one, with $w$ real-valued. It is easy to check that the extremal values for these parameters recover the field distributions used in the previous sections to study GT and dAT critical lines, respectively.

\begin{figure*}[!t]
	\centering
	\includegraphics[width=\textwidth]{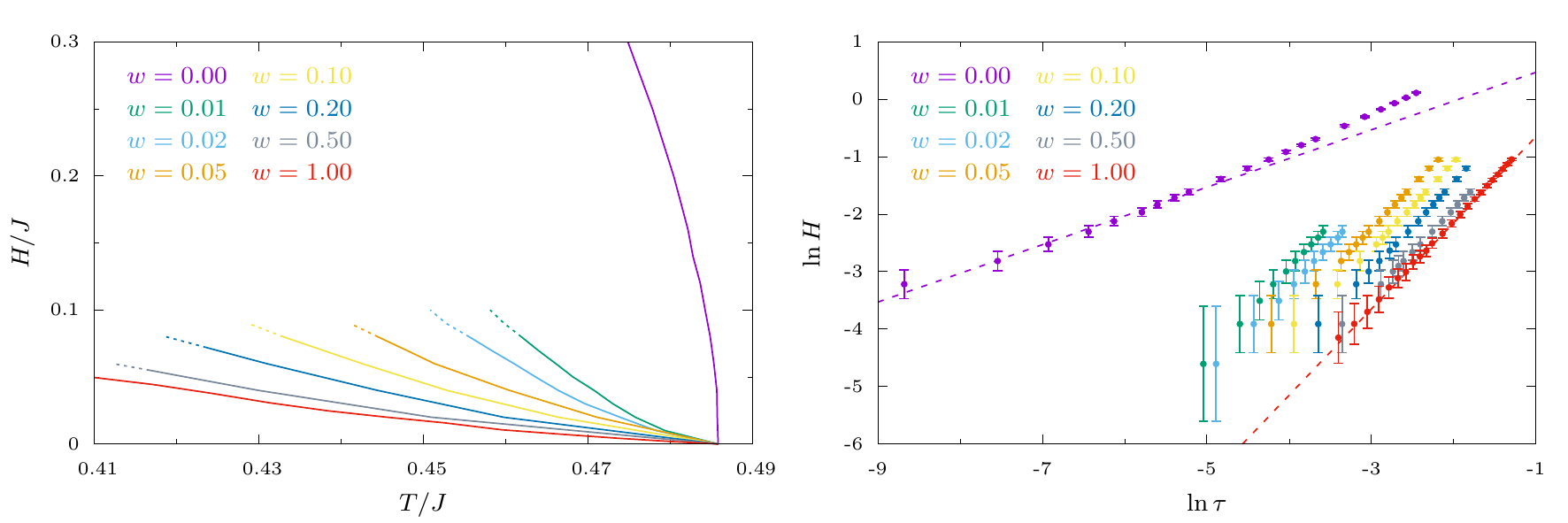}
	\caption{Critical lines in the $(T,H)$ plane for the spin glass XY model on a $C=3$ RRG with field directions $\phi=2\pi\kappa/Q$ with $\kappa$ randomly extracted according to $\mathbb{P}[\kappa=n]=(1-w+w/Q)\delta_{n,0}+w/Q\sum_{i=1}^{Q-1}\delta_{n,i}$. For $w=0$ and $w=1$ we recover the GT and the dAT lines, respectively. Data in the right panel seem to suggest a dAT-like critical behavior for any $w>0$ (dashed lines have slopes $1/2$ and $3/2$, respectively).}
	\label{fig:secondClass}
\end{figure*}

In Fig.~\ref{fig:firstClass} we plot the critical lines obtained for $C=3$ using the first class of field distributions with different values of the parameter $Q'$. Remind that $Q'=1$ and $Q'=Q=64$ correspond respectively to GT and dAT lines.
In the left panel we see that even with the smallest nontrivial value $Q'=2$ the critical line moves sensibly: so the loss of the perfect alignment among the local directions of the external field seems to have a visible effect on the critical properties of the model.
In the right panel we study in more detail the behavior of the critical lines close to the zero-field critical point: while the extremal case $Q'=1$ follows a power law with the GT-like exponent $1/2$, for $Q'>1$ the data seem to follow the dAT-like exponent $3/2$ (dashed lines have slopes $1/2$ and $3/2$, respectively).
So, to the best of our numerical evidences, the GT-like critical behavior seems to be relegated to the singular $Q'=1$ case, where all the external fields are perfectly aligned.

Given that in the first class of distributions there is a minimal perturbation $O(1/Q)$ to the GT-like distribution, we study now the second class of field distributions, where the intensity of the perturbation with respect to the $\delta(\phi)$ distribution is given by the continuous parameter $w$.
In Fig.~\ref{fig:secondClass} we show the results obtained with the second class of interpolating functions. Also in this case we notice that even the smallest $w=0.01$ perturbation produces a sensible effect on the critical line, that changes from a GT-like shape to a dAT-like shape (see left panel). Moreover, the analysis in the vicinity of the zero-field critical point shown in the right panel strongly suggests that for any $w>0$ the critical lines have the exponent $3/2$ corresponding to the dAT line. If any GT-like behavior is eventually present it would show up only in a region of extremely small values of $\tau$ and $H$ which is not easily accessible numerically.

These observations are coherent with the claim that a GT-like transition is possible if and only if the model admits the solution $q_{\perp}=0$, whose loss of stability just defines the GT line. Since this is possible only in the case of a homogeneous field over the whole system, hence our claim is that any infinitesimal perturbation to the homogeneous distribution of the field would make the GT transition disappear in favour of the dAT transition, so greatly enhancing the stability of the paramagnetic solution. GT transition is then a singular case, while the most generic and robust mechanism of RSB for a vector spin glass in a field is hence represented by the dAT transition.

\section{Conclusions}
\label{sec:concl}

We have shown how to compute critical lines in the $(T,H)$ plane for a XY spin glass model on a random regular graph. We have used different distributions of the field direction in order to probe different critical behaviors. We have identified GT-like and dAT-like critical behaviors. The corresponding critical lines in the $(T,H)$ plane are similar to the fully connected case in the vicinity of the zero-field critical point, $H_{\text{GT}}\propto \tau^{1/2}$ and $H_{\text{dAT}} \propto \tau^{3/2}$, but differ sensibly at low temperatures (as in the Ising case \cite{ParisiEtAl2014}).

We have then shown how different are the local fluctuations that become critical in the two cases: they are strongly orthogonal to the local field in the GT case, while they are mostly longitudinal in the dAT case.

Finally, we have analyzed intermediate cases, where the fields are neither fully aligned nor completely random in direction. These cases have never been studied before, to the best of our knowledge.
The comparison of the results obtained with two classes of field direction distributions interpolating between the delta function in $\phi=0$ and the flat distribution $\phi\in[0,2\pi)$ seems to suggest that the GT-like critical behavior is very unstable with respect to any small perturbation. In practice we only observe the dAT-like critical behavior for any field distribution that deviates (even by a tiny amount of order $10^{-2}$) from the situation with all the external fields perfectly aligned.

The overall picture resulting from our analysis is that the GT-like critical behavior can take place only if all the fields are perfectly aligned, while the dAT-like behavior is much more robust and generic, representing the mechanism through which replica symmetry typically breaks for vector spin glass models in a field.

\begin{acknowledgments}
	The authors thank Giorgio Parisi for useful discussions.
	This research has been supported by the European Research Council (ERC) under the European Unions Horizon 2020 research and innovation programme (grant agreement No [694925]).
\end{acknowledgments}

\appendix

\section{GT and dAT lines in the large connectivity limit}
\label{app:SK_limit}

In the main text we refer to the computation of the GT and dAT lines in an external field (respectively homogeneous over the whole system or randomly oriented on each site) on fully connected graphs, i.e.\ in the SK model, that has been already accomplished via the standard replica approach in the literature. In this appendix we want to pursue a twofold goal: first of all, we recall the replica results, explicitly writing them for the $m=2$ case, i.e.\ the XY model; then, we obtain the saddle-point equations for the fully connected XY model in a more straightforward and simpler way, via the large-connectivity limit of the belief-propagation equations; finally, we prove the equivalence of the two approaches, so providing a more direct physical interpretation of the quantities appearing in the replica computations.

\subsection{Replica results}

\subsubsection*{The uniform-field case}

On the fully connected geometry, the replica trick~\cite{Book_MezardEtAl1987} allows to succesfully solve the spin glass vector model in an external magnetic field, leading to the detection of the GT line or the dAT line depending on the distribution of local directions of the field. In particular, in the homogeneous case, the RS computation has been carried out for generic values of the number $m$ of spin components by Gabay and Toulouse~\cite{GabayToulouse1981} and later by Cragg, Sherrington and Gabay~\cite{CraggEtAl1982}. For the XY model, the saddle-point equations describing the paramagnetic solution ($q_{\perp}=0$) read:
\begin{subequations}
	\begin{equation}
		q_{\parallel} = \int_{-\infty}^{\infty}\frac{dz}{\sqrt{2\pi}}e^{-z^2/2}\biggl(\frac{P_{01}}{P_{00}}\biggr)^2
	\end{equation}
	\begin{equation}
		x = -1+\int_{-\infty}^{\infty}\frac{dz}{\sqrt{2\pi}}e^{-z^2/2}\biggl(2-\frac{P_{20}}{P_{00}}\biggr)
	\end{equation}
\end{subequations}
with $x$ known as the \textit{quadrupolar parameter}. Notice that, as usual in the replica computations, the spins are not taken with unit norm, rather $\sum_{\mu=1}^m \sigma^2_{\mu}=m$ (i.e.\ $2$ for the XY model). Functions $P_{\mu\nu}$'s appearing inside the Gaussian averages are then defined for the $m=2$ case as follows:
\begin{equation}
\begin{aligned}
	P_{\mu\nu}=\int_{-\sqrt{2}}^{\sqrt{2}}dS\,&e^{\,\beta(z\sqrt{q_{\parallel}}+H)S+(\beta^2/2)(2x-q_{\parallel}) S^2}\\
	&\times(2-S^2)^{(\mu-1)/2}S^{\nu}
	\label{eq:Bessel_like_functions_replica}
\end{aligned}
\end{equation}
Such solution is stable until the following condition is satisfied:
\begin{equation}
	\beta^2\int_{-\infty}^{\infty}\frac{dz}{\sqrt{2\pi}}e^{-z^2/2}\biggl(\frac{P_{20}}{P_{00}}\biggr)^2=1
	\label{eq:stab_cond_GT_replica}
\end{equation}
Then, below the corresponding critical line, the stable solution is characterized by a nonvanishing transverse overlap $q_{\perp}$, together with a breaking of the replica symmetry. However, here we restrict ourselves to the RS analysis, being enough for our purposes.

The small-field expansion of the condition in Eq.~(\ref{eq:stab_cond_GT_replica}) yields the well-known $1/2$ exponent of the GT line:
\begin{equation}
	H_c \propto \tau^{1/2}
\end{equation}
while in the opposite limit we have an exponential divergence of the inverse critical temperature:
\begin{equation}
	\beta_c \propto e^{\,H^2/4}
\end{equation}

Since these equations are obtained by using spins with norm $m=2$, it is useful to rewrite them for spins with unit norm, accordingly to all the computations of the main text:
\begin{equation}
	S \rightarrow \tilde{S} \equiv S/\sqrt{2}
\end{equation}
Coherently with this choice, a dimensional analysis in the Hamiltonian leads to the corresponding rescaling of temperature and field:
\begin{equation}
	\beta \rightarrow \tilde{\beta} \equiv 2\beta \quad , \quad H \rightarrow \tilde{H} \equiv H/\sqrt{2}
\end{equation}
Bessel-like functions~(\ref{eq:Bessel_like_functions_replica}) then become:
\begin{equation}
\begin{aligned}
	P_{\mu\nu}&=2^{(\mu+\nu)/2}\int_{-1}^{1}d\tilde{S}\,e^{\,\tilde{\beta}(z\sqrt{\tilde{q}_{\parallel}}+\tilde{H})\tilde{S}+(\tilde{\beta}^2/2)(\tilde{x}-\tilde{q}_{\parallel})\tilde{S}^2}\\
	&\qquad\qquad\qquad\times(1-\tilde{S}^2)^{(\mu-1)/2}\tilde{S}^{\nu}\\
	&\equiv 2^{(\mu+\nu)/2}\tilde{P}_{\mu\nu}
\end{aligned}
\end{equation}
so that we finally get also the proper rescaling of the longitudinal overlap~$q_{\parallel}$ and of the quadrupolar parameter~$x$ moving between the two normalizations:
\begin{equation}
	\tilde{x} \equiv x \quad , \quad \tilde{q}_{\parallel} \equiv q_{\parallel}/2
\end{equation}

Looking at the definition of the Bessel-like functions~(\ref{eq:Bessel_like_functions_replica}), it is easy to recognize $\tilde{S}$ as the projection of the unit spin $\tilde{\boldsymbol{S}}$ onto the $\hat{x}$ axis, namely $\tilde{S}=\cos{\theta}$. Moving to the angular variable $\theta$, then, we get:
\begin{equation}
\begin{aligned}
	\tilde{P}_{\mu\nu}&=\int_{0}^{2\pi}d\theta\,e^{\,\tilde{\beta}(z\sqrt{\tilde{q}_{\parallel}}+\tilde{H})\cos{\theta}+(\tilde{\beta}^2/2)(\tilde{x}-\tilde{q}_{\parallel})\cos^2{\theta}}\\
	&\qquad\qquad\times\sin^{\mu}{\theta}\cos^{\nu}{\theta}
\end{aligned}
\end{equation}
namely we get a sort of average of the quantity $\sin^{\mu}{\theta}\cos^{\nu}{\theta}$ over $\theta\in[0,2\pi]$ via the exponential measure $\exp{[\tilde{\beta}(z\sqrt{\tilde{q}_{\parallel}}+\tilde{H})\cos{\theta}+(\tilde{\beta}^2/2)(\tilde{x}-\tilde{q}_{\parallel})\cos^2{\theta}]}$. More concretely, we can introduce the following short-hand notation for such (normalized) angular averages:
\begin{equation}
	\braket{\sin^{\mu}(\theta)\cos^{\mu}(\theta)} \equiv \frac{\tilde{P}_{\mu\nu}}{\tilde{P}_{00}}
\end{equation}

In this way, one can easily recognize the physical meaning of the longitudinal overlap~$q_{\parallel}$: it represents the Gaussian average of the square average magnetization along the field direction
\begin{equation}
\begin{aligned}
	\tilde{q}_{\parallel}&=\int_{-\infty}^{\infty}\frac{dz}{\sqrt{2\pi}}e^{-z^2/2}\biggl(\frac{\tilde{P}_{01}}{\tilde{P}_{00}}\biggr)^2\\
	&\equiv\mathbb{E}_z[\braket{\cos{\theta}}^2]
\end{aligned}
\end{equation}
where $\mathbb{E}_z[\cdot]$ is indeed the expectation value over the Gaussian variable $z$. In the same manner, the quadrupolar parameter $x$ can be easily expressed in terms of angular averages:
\begin{equation}
\begin{aligned}
	\tilde{x} &= 2\int_{-\infty}^{\infty}\frac{dz}{\sqrt{2\pi}}e^{-z^2/2}\biggl(1-\frac{\tilde{P}_{20}}{\tilde{P}_{00}}\biggr)-1\\
	&\equiv 2\mathbb{E}_z[\braket{\cos^2{\theta}}]-1\\
	&=\mathbb{E}_z[\braket{\cos{2\theta}}]
\end{aligned}
\end{equation}
as well as the transverse overlap~$\tilde{q}_{\perp}$, representing the quadratic fluctuations in the direction transverse to the field:
\begin{equation}
	\tilde{q}_{\perp}=\mathbb{E}_z[\braket{\sin{\theta}}^2]
\end{equation}
hence vanishing in the paramagnetic phase.

Under this light, the replica saddle-point equations in the RS ansatz acquire a clear physical meaning: as long as the solution is paramagnetic, all the marginals are polarized in the direction of the field, with no freezing in the transverse direction. In the cold phase, instead, the marginals acquire incoherent transverse components, that result in a $\tilde{q}_{\perp}$ different from zero. Consequently, in this latter case, a further term proportional to $\sqrt{\tilde{q}_{\perp}}\sin{\theta}$ should be added in the exponential measure appearing in the definition of $\tilde{P}_{\mu\nu}$'s. In addition, notice that the three parameters $\tilde{q}_{\parallel}$, $\tilde{q}_{\perp}$ and $\tilde{x}$ are enough to describe both the phases --- still in the RS ansatz ---, since the candidate for a fourth parameter, $\mathbb{E}_z[\braket{\sin{2\theta}}]$, can be expressed in terms of the other ones due to the constraint on the spin normalization.

Finally, the stability condition~(\ref{eq:stab_cond_GT_replica}) becomes in the unit-norm frame:
\begin{equation}
	\tilde{\beta}^2\int_{-\infty}^{\infty}\frac{dz}{\sqrt{2\pi}}e^{-z^2/2}\biggl(\frac{\tilde{P}_{20}}{\tilde{P}_{00}}\biggr)^2=1
\end{equation}
namely, in terms of the angular variable $\theta$:
\begin{equation}
	\tilde{\beta}^2\,\mathbb{E}_z[\braket{\sin^2{\theta}}^2]=1
\end{equation}
which is nothing but the marginality condition for the growth rate of $\tilde{q}_{\perp}$, as it can be shown by expanding around the vanishing solution $\tilde{q}_{\perp}=0$. Such marginality condition will be even clearer when analyzing the large-$C$ limit of the cavity equations.

\begin{figure}[!t]
	\centering
	\includegraphics[width=\columnwidth]{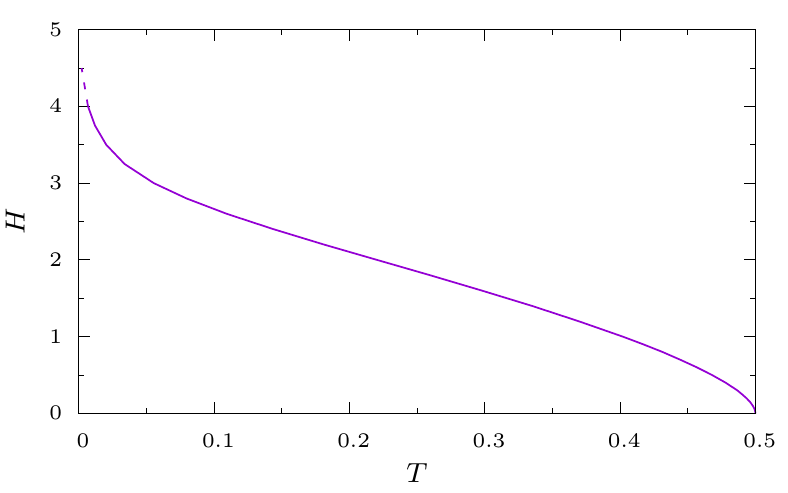}
	\caption{The GT line computed via the replica approach in the unit-norm frame.}
	\label{fig:GT_line_replica}
\end{figure}

The corresponding critical line is reported in Fig.~\ref{fig:GT_line_replica}, with the axes rescaled according to the unit-norm choice for the spins. One could easily recognize the square-root singularity close to the zero-field axis and the exponential divergence close to the zero-temperature axis.

\subsubsection*{The Gaussian-field case}

At variance, the diametrically opposite case is represented by a randomly oriented field with a flat distribution over the local directions of the field. In particular, since in replica computations one usually deals with Gaussian-distributed couplings, it is comfortable to introduce a Gaussian-distributed field as well, so that Gaussian integrals can be straightforwardly performed. Following Sharma and Young~\cite{SharmaYoung2010}, we consider each component of the field $\boldsymbol{H}$ as independently distributed according to a Gaussian of zero mean and variance $\sigma^2_H$:
\begin{equation}
	H_{\mu} \sim \mathcal{N}(0,\sigma^2_H)
	\label{eq:Gauss_field}
\end{equation}
Hence, the rotational invariance $\mathrm{O}(m)$ is restored, corresponding to a unique order parameter $q$ in the RS frame, self-consistently given --- for the XY model --- by the following equation:
\begin{equation}
	q = \int_0^{\infty} d\rho \, \rho \, e^{-\rho^2/2}\biggl[\frac{I_1(\Delta_{\mathcal{G}})}{I_0(\Delta_{\mathcal{G}})}\biggr]^2
\end{equation}
with $\Delta_{\mathcal{G}}$ containing $q$ itself, $\beta$ and the variance of the external field:
\begin{equation}
	\Delta_{\mathcal{G}} \equiv \sqrt{2}\beta\sqrt{q+\sigma^2_H}\,\rho
	\label{eq:def_Delta_Gauss}
\end{equation}

Another consequence of the rotational invariance is the absence of the quadrupolar parameter $x$, indeed being related to the breaking of the $\mathrm{O}(2)$ symmetry.

Finally, the stability of the paramagnetic RS solution can be studied via the usual techniques from the Hessian in the replica space~\cite{Book_MezardEtAl1987, SharmaYoung2010}, obtaining the following marginality condition:
\begin{equation}
	\beta^2\chi_0 = 1
	\label{eq:margCond_dAT_GaussField}
\end{equation}
with $\chi_0$ given by:
\begin{eqnarray*}
\chi_0 = 2\int_0^{\infty} d\rho \, \rho \, e^{-\rho^2/2} \biggl[ 2\frac{I^2_1(\Delta_{\mathcal{G}})}{\Delta^2_{\mathcal{G}}\,I^2_0(\Delta_{\mathcal{G}})} + 2\frac{I_1(\Delta_{\mathcal{G}})\,I_2(\Delta_{\mathcal{G}})}{\Delta_{\mathcal{G}}\,I^2_0(\Delta_{\mathcal{G}})}&&\\
+\frac{I^2_2(\Delta_{\mathcal{G}})}{I^2_0(\Delta_{\mathcal{G}})} - 2\frac{I^3_1(\Delta_{\mathcal{G}})}{\Delta_{\mathcal{G}}\,I^3_0(\Delta_{\mathcal{G}})}
- 2\frac{I^2_1(\Delta_{\mathcal{G}})\,I_2(\Delta_{\mathcal{G}})}{I^3_0(\Delta_{\mathcal{G}})} + \frac{I^4_1(\Delta_{\mathcal{G}})}{I^4_0(\Delta_{\mathcal{G}})} \biggr]&&
\end{eqnarray*}

As usual, it is easy to map these equations into the corresponding ones for the unit spins. Indeed, we already know the rescaling of $\beta$ ($\tilde{\beta}=2\beta$) and $q$ ($\tilde{q}=q/2$); then, $\sigma_H$ should rescale exactly as $H$:
\begin{equation}
	\tilde{\sigma}_H = \sigma_H / \sqrt{2}
\end{equation}
and finally we get the proper rescaling also for $\Delta_{\mathcal{G}}$:
\begin{equation}
	\tilde{\Delta}_{\mathcal{G}} \equiv \tilde{\beta}\sqrt{\tilde{q}+\tilde{\sigma}^2_H}\rho
\end{equation}
i.e.\ $\tilde{\Delta}_{\mathcal{G}}=\Delta_{\mathcal{G}}$. The equation for $\tilde{q}$, then, reads:
\begin{equation}
	\tilde{q} = \frac{1}{2}\int_0^{\infty} d\rho \, \rho \, e^{-\rho^2/2}\biggl[\frac{I_1(\tilde{\Delta}_{\mathcal{G}})}{I_0(\tilde{\Delta}_{\mathcal{G}})}\biggr]^2
\end{equation}
and finally the marginality condition~(\ref{eq:margCond_dAT_GaussField}) becomes:
\begin{equation}
	\tilde{\beta^2}\tilde{\chi}_0=1
\end{equation}
where we have defined $\tilde{\chi}_0 \equiv \chi_0/4$.

\subsubsection*{The random-field case with constant intensity $H$}

Since in the main text we have not used a Gaussian-distributed field, rather a randomly oriented field with a constant intensity $H$, we would like here to obtain the corresponding dAT line in the fully connected limit, since in principle it could be different from the one recalled above. To this aim, it is enough to look at the definition of the quantity $\Delta_{\mathcal{G}}$ in Eq.~(\ref{eq:def_Delta_Gauss}): $q+\sigma^2_H$ indeed represents the \textit{total} variance of the Gaussian field acting on each site, composed by an ``intrinsic'' variance~$q$ (due to the contributions from the neighbours) and an ``external'' contribution~$\sigma^2_H$ (due to the proper magnetic field~$\boldsymbol{H}$).

Hence, in the case of a randomly oriented field with constant intensity, we just get rid of $\sigma_H$. However, a counterpart should put into the first moment of the external field: in more detail, a bias $H\cos{\phi}$ should be considered along the~$\hat{x}$ direction and $H\sin{\phi}$ along the~$\hat{y}$ direction, forcing us to move from polar coordinates $(\rho,\vartheta)$ to Cartesian coordinates $(z_x,z_y)$. Finally, we must average over $\phi$ via the flat distribution $1/2\pi$. The argument of Bessel functions consequently changes from $\Delta_{\mathcal{G}}$ to $\Delta_{\mathcal{R}}$ so defined:
\begin{equation}
	\Delta_{\mathcal{R}} \equiv \sqrt{2}\beta\sqrt{(H\cos{\phi}+z_x\sqrt{q})^2+(H\sin{\phi}+z_y\sqrt{q})^2}
\end{equation}
and finally, via a gauge transformation over the local direction $\phi$ of the external field --- since the sum of all the messages coming from the neighbours is $\mathrm{O}(2)$ symmetric as well --- we can get rid of the average over $\phi$, getting the following definition for $\Delta_{\mathcal{R}}$:
\begin{equation}
	\Delta_{\mathcal{R}} \equiv \sqrt{2}\beta\sqrt{(H+z_x\sqrt{q})^2+(z_y\sqrt{q})^2}
\end{equation}
and the following self-consistency equation for $q$:
\begin{equation}
\begin{aligned}
	q &= \int_{-\infty}^{\infty}dz_x\,dz_y\,\frac{e^{-(z^2_x+z^2_y)/2}}{2\pi}\biggl[\frac{I_1(\Delta_{\mathcal{R}})}{I_0(\Delta_{\mathcal{R}})}\biggr]^2\\
	&\equiv \mathbb{E}_{\boldsymbol{z}}\biggl[\frac{I^2_1(\Delta_{\mathcal{R}})}{I^2_0(\Delta_{\mathcal{R}})}\biggr]
	\label{eq:q_dAT_GaussField}
\end{aligned}
\end{equation}
with $\mathbb{E}_{\boldsymbol{z}}[\cdot]$ being a short-hand notation for the Gaussian average over $\boldsymbol{z}=(z_x,z_y)$.

Finally, also the marginality condition~(\ref{eq:margCond_dAT_GaussField}) reads formally the same, i.e.\ $\beta^2\chi_0=1$, once coherently moved from $\Delta_{\mathcal{G}}$ to $\Delta_{\mathcal{R}}$:
\begin{eqnarray*}
\chi_0 = 2\,\mathbb{E}_{\boldsymbol{z}}\biggl[ 2\frac{I^2_1(\Delta_{\mathcal{R}})}{\Delta^2_{\mathcal{R}}\,I^2_0(\Delta_{\mathcal{R}})} + 2\frac{I_1(\Delta_{\mathcal{R}})\,I_2(\Delta_{\mathcal{R}})}{\Delta_{\mathcal{R}}\,I^2_0(\Delta_{\mathcal{R}})} + \frac{I^2_2(\Delta_{\mathcal{R}})}{I^2_0(\Delta_{\mathcal{R}})}&&\\
- 2\frac{I^3_1(\Delta_{\mathcal{R}})}{\Delta_{\mathcal{R}}\,I^3_0(\Delta_{\mathcal{R}})}- 2\frac{I^2_1(\Delta_{\mathcal{R}})\,I_2(\Delta_{\mathcal{R}})}{I^3_0(\Delta_{\mathcal{R}})} + \frac{I^4_1(\Delta_{\mathcal{R}})}{I^4_0(\Delta_{\mathcal{R}})} \biggr]&&
\end{eqnarray*}

Also in this case, the mapping to the unit-norm frame is quite straightforward, being:
\begin{equation}
	\tilde{\beta} = 2\beta \quad , \quad \tilde{q} = q/2 \quad , \quad \tilde{H} = H/\sqrt{2}
\end{equation}
and from them the definition of $\tilde{\Delta}_{\mathcal{R}}$:
\begin{equation}
	\tilde{\Delta}_{\mathcal{R}} \equiv \tilde{\beta}\sqrt{(\tilde{H}+\sqrt{\tilde{q}}\,z_x)^2+(\sqrt{\tilde{q}}\,z_y)^2}
\end{equation}
so that $\tilde{\Delta}_{\mathcal{R}}=\Delta_{\mathcal{R}}$. The equation~(\ref{eq:q_dAT_GaussField}) for $q$ then becomes:
\begin{equation}
	\tilde{q} = \frac{1}{2}\mathbb{E}_{\boldsymbol{z}}\biggl[\frac{I^2_1(\tilde{\Delta}_{\mathcal{R}})}{I^2_0(\tilde{\Delta}_{\mathcal{R}})}\biggr]
\end{equation}
and finally we get again that the marginality condition reads $\tilde{\beta}^2\tilde{\chi}_0=1$ with $\tilde{\chi}_0\equiv\chi_0/4$.

\begin{figure}[!t]
	\centering
	\includegraphics[width=\columnwidth]{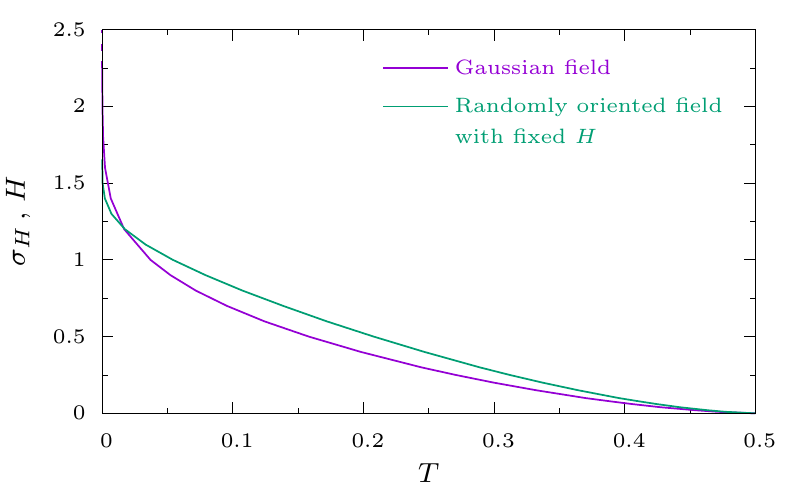}
	\caption{The dAT line computed via the replica approach in the unit-norm frame, obtained when using respectively a Gaussian distribution for the field components (purple curve) and a randomly oriented field with constant intensity (green curve).}
	\label{fig:dAT_line_replica}
\end{figure}

At this point, we can compare the two choices for the local distribution of the external field. As anticipated, they yield different shapes of the dAT line in the $T$ vs $H$ plane, as it can be appreciated in Fig.~\ref{fig:dAT_line_replica}. First of all, they have the same behaviour in the small-field limit, namely $H \propto \tau^{3/2}$, but a different coefficient in front of such term. This is due to the fact that in the Gaussian-field case the stability of the paramagnetic phase is enhanced by the rare presence of some exceptionally intense field, while this phenomenon is not possible in the case of the random field with fixed modulus $H$.

Secondly, and most importantly, a rather different behaviour when approaching the zero-temperature limit. Indeed, in the $\beta\to\infty$ limit, both $\Delta_{\mathcal{G}}$ and $\Delta_{\mathcal{R}}$ diverge linearly with~$\beta$, so that $\chi_0$ can be expanded in power series of $1/\Delta$. The first nonvanishing contribution of $\chi_0$ is given by the term $2/\Delta^2 \propto \beta^{-2}$, as expected. So when substituting into the marginality condition $\beta^2\chi_0=1$ we get (in the $m=2$-norm setting, so to match with the literature results):
\begin{equation}
	\frac{2\beta^2}{\Delta^2_{\mathcal{G},\mathcal{R}}}=1
	\label{eq:margCond_dAT_zeroTemp}
\end{equation}
where $q$ within $\Delta_{\mathcal{G},\mathcal{R}}$ can be already set equal to $1$, so neglecting higher-order corrections.

In the Gaussian case, such condition explicitly becomes:
\begin{equation}
\begin{aligned}
	&\int_{0}^{\infty} d\rho \, \rho \, \frac{e^{-\rho^2/2}}{(1+\sigma^2_H)\rho^2} = 1\\
	&\qquad\qquad \Rightarrow \quad \frac{1}{1+\sigma^2_H}\int_{0}^{\infty} d\rho \, \frac{e^{-\rho^2/2}}{\rho} = 1
	\label{eq:margCond_dAT_zeroTemp_GaussField}
\end{aligned}
\end{equation}
which can not be satisfied for any finite value of $\sigma_H$, being the integral in $\rho$ divergent. Hence, according to the prediction by Sharma and Young~\cite{SharmaYoung2010}, the dAT line approaches the zero-temperature axis only asymptotically when considering a Gaussian distribution for the field components in the $m=2$ case, while it touches the $T=0$ axis at a finite value of $\sigma_H$ for $m \geqslant 3$.

Analogously, in the case of a randomly oriented field with constant intensity $H$, the marginality condition at $T=0$ reads:
\begin{equation}
	\int_{-\infty}^{\infty} dz_x \, dz_y \, \frac{e^{-(z^2_x+z^2_y)/2}}{2\pi}\,\frac{1}{(H+z_x)^2+z_y^2} = 1
	\label{eq:margCond_dAT_zeroTemp_GaugeGlassField}
\end{equation}
from which, via some manipulations, we get again a divergent integral on the left hand side of the marginality condition
\begin{equation}
	\int_{1}^{\infty} d\rho \, \frac{e^{-H^2(\rho-1)/2\rho}}{2\rho} = \infty \quad \forall\,H
\end{equation}
implying a divergent value of the critical field $H$ in the $T\to 0$ limit.

The divergence of the integrals in the two cases can be then exploited in order to check the rate at which the critical variance $\sigma_H$ and the critical field $H$, respectively, diverge in the $\beta\to\infty$ limit. Indeed, from the inspection of Fig.~\ref{fig:dAT_line_replica} it is clear that they approach the $T=0$ axis in a rather different manner, with the curve $H(T)$ converging faster than the curve $\sigma_H(T)$. To this aim, let us define the function $f(\Delta)$ such that its Gaussian average over $\boldsymbol{z}$ gives $\chi_0$:
\begin{equation}
	f(\Delta): \quad \chi_0 \equiv \mathbb{E}_{\boldsymbol{z}}[f(\Delta)]
\end{equation}
Moreover, we already know its behaviour in the two opposite regimes of small- and large-argument limits, valid in both cases of a Gaussian field and a randomly oriented field with constant intensity:
\begin{equation}
	f(\Delta=0)=1 \quad , \quad f(\Delta\gg 1) \simeq \frac{2}{\Delta^2}
\end{equation}

\begin{figure}[!t]
	\centering
	\includegraphics[width=\columnwidth]{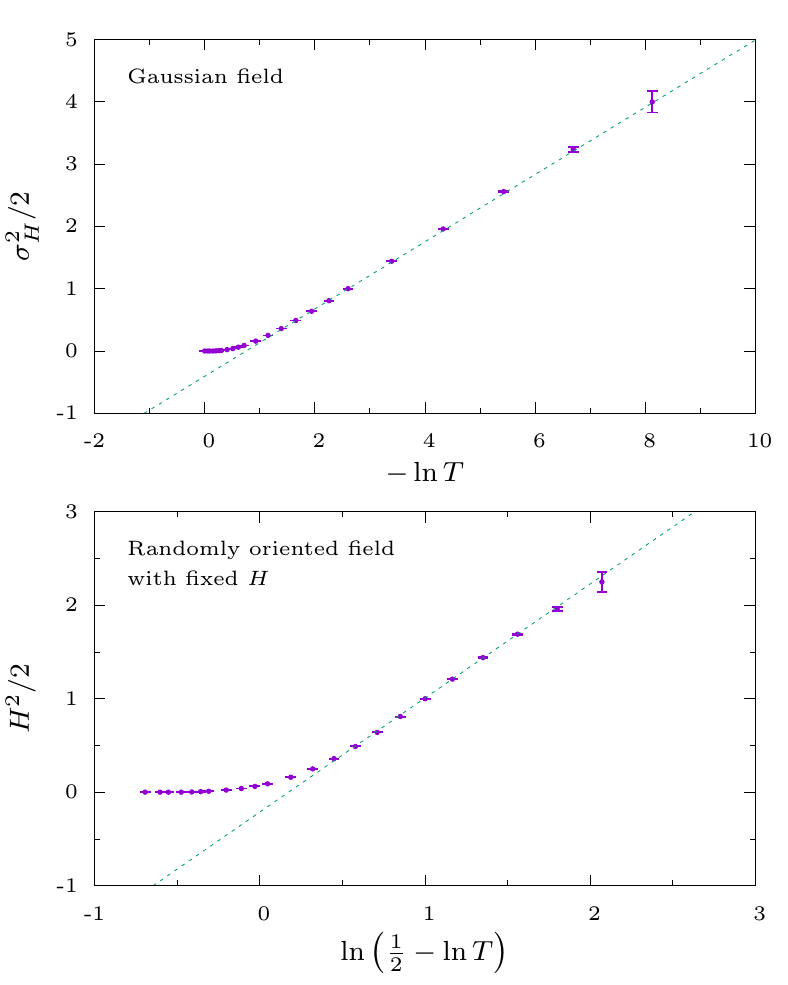}
	\caption{Convergence to zero of the critical temperature along the dAT line when increasing the field variance for the Gaussian case (upper panel) or the field strength for the randomly oriented case with fixed $H$ (lower panel). The linear trend for large values of $\sigma_H$ and $H$ confirms the analytic results~(\ref{eq:dAT_stab_largeSigmaH_GaussField}) and~(\ref{eq:dAT_stab_largeH_GaugeGlassField}), respectively. Axes scale refer to the $m=2$-norm choice for the spins, while error bars are due to the numeric precision used in the computation.}
	\label{fig:dAT_line_FC_TempCloseToZero}
\end{figure}

So let us now analyze the condition $\beta^2\chi_0=1$ for large but finite values of $\beta$ in the Gaussian case. We have:
\begin{equation}
	\beta^2\int_0^{\infty} d\rho \, \rho \, e^{-\rho^2/2} f(\Delta_{\mathcal{G}}) = 1
\end{equation}
The argument $\Delta_{\mathcal{G}}$ becomes $\sqrt{2}\beta\sigma_H\rho$, being $q$ negligible with respect to the critical value of $\sigma^2_H$ in the low-temperature limit. Then, the divergence of the integral in the $\beta\to\infty$ limit can be controlled by dividing the integration domain in two regions, respectively $\mathcal{A} \equiv [0,\epsilon]$ and $\mathcal{B} \equiv [\epsilon,\infty)$. In the region $\mathcal{A}$, we get that the Gaussian weight can be neglected; then, we perform a change of variables, $\sqrt{2}\beta\sigma_H\rho \equiv x$:
\begin{equation}
\begin{aligned}
	&\beta^2\int_{\mathcal{A}} d\rho \, \rho \, e^{-\rho^2/2} f(\sqrt{2}\beta\sigma_H\rho)\\
	&\qquad\simeq \beta^2\int_{\mathcal{A}} d\rho \, \rho \, f(\sqrt{2}\beta\sigma_H\rho)\\
	&\qquad\simeq \frac{1}{2\sigma^2_H}\int_{0}^{\sqrt{2}\beta\sigma_H\epsilon} dx \, x \, f(x)\\
	&\qquad\simeq \frac{1}{2}\beta^2\epsilon^2
\end{aligned}
\end{equation}
having exploited the limit $\lim_{x\to 0}f(x)=1$. Since the integral on the region $\mathcal{A}$ has to be finite in the $\beta\to\infty$ limit, then $\epsilon$ should scale as the inverse power of it:
\begin{equation}
	\epsilon \sim \frac{1}{\beta}
\end{equation}
Let us now move to the integration over the $\mathcal{B}$ region. In this region, $f$ can be approximated with the first term of its expansion for large arguments, giving:
\begin{equation}
\begin{aligned}
	&\beta^2\int_{\mathcal{B}} d\rho \, \rho \, e^{-\rho^2/2} f(\sqrt{2}\beta\sigma_H\rho)\\
	&\qquad\simeq \beta^2\int_{\mathcal{B}} d\rho \, \rho \, e^{-\rho^2/2} \, \frac{2}{2\beta^2\sigma^2_H\rho^2}\\
	&\qquad\simeq \frac{1}{\sigma^2_H}\int_{\epsilon}^{\infty} d\rho \, \frac{e^{-\rho^2/2}}{\rho}\\
	&\qquad\simeq -\frac{1}{\sigma^2_H}\ln{\epsilon}
\end{aligned}
\end{equation}
Finally, when taking $\epsilon \sim 1/\beta$ for both the contributions, we get that the marginality condition reads:
\begin{equation}
	\frac{1}{2}+\frac{1}{\sigma^2_H}\ln{\beta} = 1
\end{equation}
from which the scaling of the inverse critical temperature $\beta_c$ with $\sigma_H$:
\begin{equation}
	\beta_c \propto e^{\,\sigma^2_H/2}
	\label{eq:dAT_stab_largeSigmaH_GaussField}
\end{equation}
that can be also appreciated in the upper panel of Fig.~\ref{fig:dAT_line_FC_TempCloseToZero}.

An analogous reasoning leads to the prediction of the growth of $\beta$ with $H$ along the dAT line in the random-field case with fixed $H$. Indeed, we have that the integral in the marginality condition
\begin{equation}
	\beta^2\int_{-\infty}^{\infty} dz_x \, dz_y \, \frac{e^{-(z^2_x+z^2_y)/2}}{2\pi}\,f(\Delta_{\mathcal{R}}) = 1
\end{equation}
can be again divided in two regions, $\mathcal{A}$ and $\mathcal{B}$, where $\mathcal{A}$ is the disk of radius $\epsilon$ centered around the point $(-H,0)$ and $\mathcal{B}$ is the remaining portion of the $(z_x,z_y)$ plane. As before, in the region $\mathcal{A}$ the Gaussian weight can be considered constant; then, we move to polar coordinates and perform the change of coordinates $x \equiv \sqrt{2}\beta\rho$
\begin{equation}
\begin{aligned}
	&\beta^2\int_{\mathcal{A}} dz_x \, dz_y \, \frac{e^{-(z^2_x+z^2_y)/2}}{2\pi}\,f(\Delta_{\mathcal{R}})\\
	&\qquad\simeq \beta^2\int_{\mathcal{A}} dz_x \, dz_y \, \frac{e^{-H^2/2}}{2\pi}\,f(\Delta_{\mathcal{R}})\\
	&\qquad\simeq \beta^2 \, e^{-H^2/2}\int_0^{\epsilon} d\rho \, \rho \, f(\sqrt{2}\beta\rho)\\
	&\qquad\simeq \frac{1}{2}e^{-H^2/2}\int_0^{\sqrt{2}\beta\epsilon} dx \, x \, f(x)\\
	&\qquad\simeq \frac{1}{2}\beta^2\epsilon^2 \, e^{-H^2/2}
\end{aligned}
\end{equation}
where again the proper rescaling of the radius $\epsilon$ of the region $\mathcal{A}$ should be as $1/\beta$ when increasing $\beta$. Then, considering the integral over the region $\mathcal{B}$, we can substitute $f$ by its large-argument expansion, and then move again to polar coordinates:
\begin{equation}
\begin{aligned}
	&\beta^2\int_{\mathcal{B}} dz_x \, dz_y \, \frac{e^{-(z^2_x+z^2_y)/2}}{2\pi}\,f(\Delta_{\mathcal{R}})\\
	&\qquad\simeq \beta^2\int_{\mathcal{B}} dz_x \, dz_y \, \frac{e^{-(z^2_x+z^2_y)/2}}{2\pi}\,\frac{2}{2\beta^2[(H+z_x)^2+z^2_y]}\\
	&\qquad\simeq e^{-H^2/2}\int_{\epsilon}^{\infty} d\rho \, \frac{e^{-\rho^2/2}}{\rho}\int_0^{2\pi}d\vartheta\,\frac{e^{H\rho\cos{\vartheta}}}{2\pi}\\
	&\qquad\simeq e^{-H^2/2}\int_{\epsilon}^{\infty} d\rho \, \frac{e^{-\rho^2/2}}{\rho}I_0(H\rho)\\
	&\qquad\simeq -e^{-H^2/2}\ln{\epsilon}
\end{aligned}
\end{equation}
So, taking again $\epsilon \sim 1/\beta$, we have:
\begin{equation}
	\frac{1}{2}e^{-H^2/2}+e^{-H^2/2}\ln{\beta} = 1
\end{equation}
from which the scaling of $\beta$ with $H$ along the dAT line in the large-field region:
\begin{equation}
	\beta_c \propto \exp{\left\{e^{H^2/2}-\frac{1}{2}\right\}}
	\label{eq:dAT_stab_largeH_GaugeGlassField}
\end{equation}
numerically confirmed by the lower panel of Fig.~\ref{fig:dAT_line_FC_TempCloseToZero}.

These computations just confirm the feeling given by Fig.~\ref{fig:dAT_line_replica} that the dAT line approaches the $T=0$ axis much more rapidly in the random-field case with respect to the Gaussian-field case, gaining an exponential factor. An analogous ``exponential speedup'' can be observed in the Ising model, where the dAT line in the case of a Gaussian field goes as $\beta \propto \sigma_H$~\cite{Bray1982}, while in the case of a field with constant intensity it goes as $\beta \propto \exp{\{H^2/2\}}$~\cite{deAlmeidaThouless1978}. The reason lies in the observation that in the case of Gaussian-distributed field, with finite probability we may observe small enough fields that make the system more unstable with respect to the case of a field with constant intensity at the same temperature $T$.

\subsection{The SK limit from the BP equations}

Even though providing a formal tool through which solve spin glass models on fully connected graphs, the replica method is often quite involved, so that the physical interpretation of what is actually happening at the critical point remains hidden. At variance, the belief-propagation method bases on a very intuitive idea, symmetries are always exploited in a clear manner and the phase transitions can be typically detected via a standard analysis of the linear stability of fixed points.

In this spirit, we would like to recover the replica results via a suitable large-$C$ expansion of the~BP~equations, that at variance have been numerically solved in the main text in the case $C=O(1)$. To this aim, it is more convenient to use the factor-graph notation~\cite{Book_MezardMontanari2009} with both $\eta$'s and $\hat{\eta}$'s cavity marginals --- though still considering just pairwise interactions ---, then rewriting them as large-deviation functions in $\beta$, as done in the zero-temperature limit~\cite{LupoRicciTersenghi2017a, Thesis_Lupo2017}:
\begin{equation}
	\eta \equiv \exp{(\beta h)} \quad , \quad \hat{\eta} \equiv \exp{(\beta u)}
\end{equation}
Moreover, in order to lighten the notation and also to generalize the result to the case $m>2$, in this section we denote each spin with the unit vector $\boldsymbol{\sigma}_i$ rather with its angular variables. We will go back to the XY case when making explicit the distribution of the local direction of the external field.

Along a given directed edge $k\to i$, we have both the variable-to-check cavity message $\eta_{k\to i}(\boldsymbol{\sigma}_k) \equiv \exp{[\beta h_{k\to i}(\boldsymbol{\sigma}_k)]}$ and the check-to-variable cavity message $\hat{\eta}_{k\to i}(\boldsymbol{\sigma}_i) \equiv \exp{[\beta u_{k\to i}(\boldsymbol{\sigma}_i)]}$, that transform into each other when encountering the interaction node:
\begin{equation}
	\hat{\eta}_{k\to i}(\boldsymbol{\sigma}_i) \cong \int{d\boldsymbol{\sigma}_k} \, e^{\,\beta J_{ik}\boldsymbol{\sigma}_i\cdot\boldsymbol{\sigma}_k}\eta_{k\to i}(\boldsymbol{\sigma}_k)
	\label{eq:BP_XY_eq_etahat}
\end{equation}
apart from a normalizing multiplicative constant, or exploiting the large-deviation formalism:
\begin{equation}
	e^{\,\beta u_{k\to i}(\boldsymbol{\sigma}_i)} \cong \int{d\boldsymbol{\sigma}_k} \, e^{\,\beta[J_{ik}\boldsymbol{\sigma}_i\cdot\boldsymbol{\sigma}_k+h_{k\to i}(\boldsymbol{\sigma}_k)]}
	\label{eq:BP_XY_eq_u}
\end{equation}
Eventually, once arrived into the node $i$, the external field acting on it (about which we do not make any assumption for the moment) and the bias given by the other neighbours allow to write the expression for the variable-to-check cavity message $\eta_{i\to j}(\boldsymbol{\sigma}_i)$ along the directed edge $i\to j$:
\begin{equation}
	\eta_{i\to j}(\boldsymbol{\sigma}_i) \cong e^{\,\beta \boldsymbol{H}_i\cdot\boldsymbol{\sigma}_i}\prod_{k\in\partial i\setminus{j}}\hat{\eta}_{k\to i}(\boldsymbol{\sigma}_i)
	\label{eq:BP_XY_eq_eta}
\end{equation}
again up to a multiplicative constant, namely:
\begin{equation}
	h_{i\to j}(\boldsymbol{\sigma}_i) \simeq \boldsymbol{H}_i\cdot\boldsymbol{\sigma}_i+\sum_{k\in\partial i\setminus{j}} u_{k\to i}(\boldsymbol{\sigma}_i)
	\label{eq:BP_XY_eq_h}
\end{equation}
up to an additive constant. If Eqs.~(\ref{eq:BP_XY_eq_etahat}) and~(\ref{eq:BP_XY_eq_eta}) are put together, one gets back the pairwise BP equations seen in the main text.

In the large-$C$ limit --- when $C$ becomes of order $N$ --- exchange couplings $J_{ij}$'s have to be taken of order $1/\sqrt{C-1} \sim 1/\sqrt{N}$; then, the compatibility function can be expanded up to the second order in $J_{ij}$:
\begin{equation}
	e^{\,\beta J_{ik}\boldsymbol{\sigma}_i\cdot\boldsymbol{\sigma}_k} \simeq 1+\beta J_{ik}\boldsymbol{\sigma}_i\cdot\boldsymbol{\sigma}_k+\frac{\beta^2}{2}J^2_{ik}(\boldsymbol{\sigma}_i\cdot\boldsymbol{\sigma}_k)^2
\end{equation}
from which, when integrating over the spin $\boldsymbol{\sigma}_k$ as in~(\ref{eq:BP_XY_eq_u}):
\begin{equation}
\begin{aligned}
	&e^{\,\beta u_{k\to i}(\boldsymbol{\sigma}_i)}\\
	&\quad\simeq 1 + \beta J_{ik}\braket{\boldsymbol{\sigma}_i\cdot\boldsymbol{\sigma}_k}_k+\frac{\beta^2}{2}J^2_{ik}\braket{(\boldsymbol{\sigma}_i\cdot\boldsymbol{\sigma}_k)^2}_k\\
	&\quad\simeq e^{\,\beta J_{ik}\braket{\boldsymbol{\sigma}_i\cdot\boldsymbol{\sigma}_k}_k+(\beta^2/2)J^2_{ik}\bigl[\braket{(\boldsymbol{\sigma}_i\cdot\boldsymbol{\sigma}_k)^2}_k-\braket{\boldsymbol{\sigma}_i\cdot\boldsymbol{\sigma}_k}^2_k\bigr]}\\
	&\quad\simeq e^{\,\beta J_{ik}\boldsymbol{\sigma}_i\cdot\braket{\boldsymbol{\sigma}_k}_k+(\beta^2/2)J^2_{ik}\boldsymbol{\sigma}_i\cdot\bigl(\braket{\boldsymbol{\sigma}_k\boldsymbol{\sigma}^{\intercal}_k}_k-\braket{\boldsymbol{\sigma}_k}_k\braket{\boldsymbol{\sigma}^{\intercal}_k}_k\bigr)\cdot\boldsymbol{\sigma}_i}
\end{aligned}
\end{equation}
where $\boldsymbol{\sigma}^{\intercal}$ is the transpose vector of $\boldsymbol{\sigma}$, and where
\begin{equation}
	\braket{(\cdot)}_k \equiv \frac{\int d\boldsymbol{\sigma}_k (\cdot) \exp{[\beta h_{k\to i}(\boldsymbol{\sigma}_k)]}}{\int d\boldsymbol{\sigma}_k \exp{[\beta h_{k\to i}(\boldsymbol{\sigma}_k)]}}
\end{equation}
so to take into account also the proper normalization constant.

At this point, we exploit the second BP equation, (\ref{eq:BP_XY_eq_h}), to compute the cavity field $h_{i\to j}(\theta_i)$:
\begin{multline}
	\beta h_{i\to j}(\boldsymbol{\sigma}_i) \simeq \beta\boldsymbol{H}_i\cdot\boldsymbol{\sigma}_i+\beta\sum_{k\in\partial i\setminus j}J_{ik}\,\boldsymbol{\sigma}_i\cdot\braket{\boldsymbol{\sigma}_k}_k\\
	\qquad\times (\beta^2/2)\sum_{k\in\partial i\setminus j}J^2_{ik}\,\boldsymbol{\sigma}_i\cdot\bigl(\braket{\boldsymbol{\sigma}_k\boldsymbol{\sigma}^{\intercal}_k}_k-\braket{\boldsymbol{\sigma}_k}_k\braket{\boldsymbol{\sigma}^{\intercal}_k}_k\bigr)\cdot\boldsymbol{\sigma}_i
\end{multline}
Since the r.\,h.\,s. also contains the $h$'s cavity fields --- hidden into the expectation values $\braket{\cdot}_k$'s ---, such set of equations can be closed by using the following ansatz, presented in Ref.~\cite{JavanmardEtAl2016}:
\begin{equation}
	\beta h(\boldsymbol{\sigma}) \equiv \beta \boldsymbol{\xi}\cdot\boldsymbol{\sigma} + \frac{\beta^2}{2}\boldsymbol{\sigma}\cdot\mathbb{C}\cdot\boldsymbol{\sigma}
	\label{eq:largeC_matrixAnsatz_generic}
\end{equation}
where $\boldsymbol{\xi}$ is a $m$-component vector and $\mathbb{C}$ is a $m \times m$ symmetric matrix. So we get a set of cavity equations for these $\boldsymbol{\xi}$'s and $\mathbb{C}$'s:
\begin{subequations}
	\begin{equation}
		\boldsymbol{\xi}_{i\to j} = \boldsymbol{H}_i + \sum_{k\in\partial i\setminus j}J_{ik}\braket{\boldsymbol{\sigma}_k}_k
		\label{eq:largeC_matrixAnsatz_xi}
	\end{equation}
	\begin{equation}
		\mathbb{C}_{i\to j} = \sum_{k\in\partial i\setminus j}J^2_{ik}\,\bigl[\braket{\boldsymbol{\sigma}_k\boldsymbol{\sigma}^{\intercal}_k}_k-\braket{\boldsymbol{\sigma}_k}_k\braket{\boldsymbol{\sigma}^\intercal_k}_k\bigr]
		\label{eq:largeC_matrixAnsatz_C}
	\end{equation}
	\label{eq:largeC_matrixAnsatz}%
\end{subequations}

Finally, since we are summing over $C=O(N)$ neighbours with the couplings that are randomly distributed with zero mean and $O(1/N)$ variance, we get for the central-limit theorem that all the sites and the directed edges behave the same. Getting rid of the edge indexes, we get that $\boldsymbol{\xi}$~is a Gaussian-distributed vector with mean $\boldsymbol{M}$ and covariance matrix $\mathbb{Q}$:
\begin{equation}
	\boldsymbol{\xi} \sim \mathcal{N}(\boldsymbol{M},\mathbb{Q})
\end{equation}
while $\mathbb{C}$ becomes a deterministic quantity, due to the system-wide average ($J^2\approx 1/N$) on the r.\,h.\,s.\ of~(\ref{eq:largeC_matrixAnsatz_C}).

Since the Hamiltonian of vector spin glass models is generally $\mathrm{O}(m)$-invariant in absence of the external field, we expect such a symmetry to be eventually broken to $\mathrm{O}(m-1)$, either spontaneously or due to the presence of the external field. Hence, there exists a suitable rotation that makes the $\mathbb{Q}$ and $\mathbb{C}$ matrices diagonal.

The exponential measure $\exp{[\beta h(\boldsymbol{\sigma})]}$ appearing in the average $\braket{\cdot}$ can be then rewritten in terms of few parameters
\begin{equation}
	\beta h(\boldsymbol{\sigma}) = \beta \sum_{\mu=1}^m \bigl(M_{\mu}+z_{\mu}\sqrt{\mathbb{Q}_{\mu\mu}}\bigr)\sigma_{\mu} + \frac{\beta^2}{2}\sum_{\mu=1}^m\mathbb{C}_{\mu\mu}\sigma^2_{\mu}
	\label{eq:largeC_exponential_ansatz}
\end{equation}
with $z_{\mu} \sim \mathcal{N}(0,1)$, leading to a set of self-consistency equations for them:
\begin{subequations}
	\begin{equation}
		M_{\mu} \equiv \mathbb{E}_{\boldsymbol{z}}[\xi_{\mu}] = H_{\mu}
		\label{eq:largeC_matrixAnsatz_parameter_M}
	\end{equation}
	\begin{equation}
		\mathbb{Q}_{\mu\mu} \equiv \mathbb{V}_{\boldsymbol{z}}[\xi_{\mu}] = \mathbb{E}_{\boldsymbol{z}}\bigl[\braket{\sigma_{\mu}}^2\bigr]
		\label{eq:largeC_matrixAnsatz_parameter_Q}
	\end{equation}
	\begin{equation}
		\mathbb{C}_{\mu\mu} = \mathbb{E}_{\boldsymbol{z}}\bigl[\braket{\sigma^2_{\mu}}-\braket{\sigma_{\mu}}^2\bigr]
		\label{eq:largeC_matrixAnsatz_parameter_C}
	\end{equation}
	\label{eq:largeC_matrixAnsatz_parameters}%
\end{subequations}
where $\mathbb{E}_{\boldsymbol{z}}[\cdot]$ refers to the expectation value with respect to the Gaussian variables $z_{\mu}$'s, while $\mathbb{V}_{\boldsymbol{z}}[\cdot]$ is the corresponding variance. Eventually, $H_{\mu}$ is the expectation value of the field along the direction $\mu$.

In the end, we further exploit the breaking of the $\mathrm{O}(m)$ rotational symmetry to --- at most --- $\mathrm{O}(m-1)$ and the normalization constraint for the spins. Consequently, assuming as $\mu=1$ --- i.e.\ the $\hat{x}$ axis --- the direction along which the symmetry is eventually broken, we can redefine the matrix $\mathbb{C}$ up to a diagonal shift, $\mathbb{C}' \equiv \mathbb{C}-\mathbb{C}_{\mu\mu}\mathbb{I}$, getting the following saddle-point equations:
\begin{subequations}
	\begin{equation}
		M_{\mu} = 
		\left\{
		\begin{aligned}
			&H_x \quad &&\text{for } \mu=1\\
			&0 \quad &&\text{for } \mu=\{2,3,\dots,m\}
		\end{aligned}
		\right.
		\label{eq:largeC_matrixAnsatz_parameter_M_new}
	\end{equation}
	\begin{equation}
		\mathbb{Q}_{\mu\mu} = 
		\left\{
		\begin{aligned}
			&\mathbb{E}_{\boldsymbol{z}}\bigl[\braket{\sigma_x}^2\bigr] \quad &&\text{for } \mu=1\\
			&\mathbb{E}_{\boldsymbol{z}}\bigl[\braket{\sigma_{\mu}}^2\bigr] \quad &&\text{for } \mu=\{2,3,\dots,m\}
		\end{aligned}
		\right.
		\label{eq:largeC_matrixAnsatz_parameter_Q_new}
	\end{equation}
	\begin{equation}
		\mathbb{C}'_{\mu\mu} = 
		\left\{
		\begin{aligned}
			&\mathbb{E}_{\boldsymbol{z}}\Bigl[\braket{\sigma^2_x}-\braket{\sigma_x}^2
			-\bigl(\braket{\sigma^2_{\mu}}-\braket{\sigma_{\mu}}^2\bigr)\Bigr] \quad\text{for } \mu=1&&\\
			&0 \qquad\qquad\qquad\qquad\qquad\;\;\text{for } \mu=\{2,3,\dots,m\}&&
		\end{aligned}
		\right.
		\label{eq:largeC_matrixAnsatz_parameter_C_new}
	\end{equation}
	\label{eq:largeC_matrixAnsatz_parameters_new}%
\end{subequations}
where, again, $H_x$ is the expectation value of the external field along the $\hat{x}$ axis, while it has zero mean along the other directions. Finally, these equations can be completely solved only once made explicit the distribution of the local direction of the field.

\subsubsection*{The uniform field case}

Let us now go back to the XY model. In the uniform-field case, assuming that the symmetry is broken along the $\hat{x}$ axis, self-consistency equation~(\ref{eq:largeC_matrixAnsatz_parameter_M_new}) becomes:
\begin{equation}
	M_x = H \quad , \quad M_y = 0
\end{equation}
and hence we can directly get rid of $\boldsymbol{M}$, by plugging $H$ into the other equations. Then, in Eq.~(\ref{eq:largeC_matrixAnsatz_parameter_Q_new}), $\mathbb{Q}_{xx}$ is surely larger than zero at any temperature --- due to the presence of the external field ---, while $\mathbb{Q}_{yy}$ is either positive or zero depending on whether the transverse symmetry is broken or not, respectively. Finally, in the quadratic term in the exponential measure $\exp{[\beta h(\boldsymbol{\sigma})]}$ --- over which perform the average $\braket{\cdot}$ ---, we are left with the only term $\mathbb{C}'_{xx}$, as explained before. The large-$C$ ansatz for $h(\theta)$ hence reads:
\begin{equation}
\begin{aligned}
	\beta h(\theta) &=\beta\bigl(H+z_x\sqrt{\mathbb{Q}_{xx}}\bigr)\cos{\theta}+\beta z_y\sqrt{\mathbb{Q}_{yy}}\sin{\theta}\\
	&\qquad + \frac{\beta^2}{2}\mathbb{C}'_{xx}\cos^2{\theta}
	\label{eq:h_ansatz_GT}
\end{aligned}
\end{equation}
In terms of the angular variable $\theta$, the self-consistency equations for $\mathbb{Q}_{xx}$, $\mathbb{Q}_{yy}$ and $\mathbb{C}'_{xx}$ read:
\begin{subequations}
	\begin{equation}
		\mathbb{Q}_{xx} = \mathbb{E}_{\boldsymbol{z}}\bigl[\braket{\cos{\theta}}^2\bigr]
		\label{eq:largeC_matrixAnsatz_GT_parameter_Qxx}
	\end{equation}
	\begin{equation}
		\mathbb{Q}_{yy} = \mathbb{E}_{\boldsymbol{z}}\bigl[\braket{\sin{\theta}}^2\bigr]
		\label{eq:largeC_matrixAnsatz_GT_parameter_Qyy}
	\end{equation}
	\begin{equation}
	\begin{aligned}
		\mathbb{C}'_{xx} &= \mathbb{E}_{\boldsymbol{z}}\bigl[\braket{\cos^2{\theta}}-\braket{\cos{\theta}}^2-\bigl(\braket{\sin^2{\theta}}-\braket{\sin{\theta}}^2\bigr)\bigr]\\
		&=\mathbb{E}_{\boldsymbol{z}}\bigl[2\braket{\cos^2{\theta}}-1-\braket{\cos{\theta}}^2+\braket{\sin{\theta}}^2\bigr]
		\label{eq:largeC_matrixAnsatz_GT_parameter_Cxx}
	\end{aligned}
	\end{equation}
	\label{eq:largeC_matrixAnsatz_GT_parameters}%
\end{subequations}

The paramagnetic solution is the one with no breaking of the transverse symmetry, namely $\mathbb{Q}_{yy}=0$. The corresponding values of $\mathbb{Q}_{xx}$ and $\mathbb{C}'_{xx}$ have then to be determined according to Eqs.~(\ref{eq:largeC_matrixAnsatz_GT_parameter_Qxx}) and~(\ref{eq:largeC_matrixAnsatz_GT_parameter_Cxx}), with the Gaussian average meant to be over the sole $z_x$ variable. Eventually, it is straightforward to obtain the stability condition for such solution, by looking at Eq.~(\ref{eq:largeC_matrixAnsatz_GT_parameter_Qyy}) and expanding the r.\,h.\,s. at the first order in $\mathbb{Q}_{yy}$:
\begin{equation}
\begin{aligned}
	\mathbb{Q}_{yy} &= \mathbb{E}_{\boldsymbol{z}}\bigl[\braket{\sin{\theta}}^2\bigr]\\
		&\simeq \mathbb{E}_{\boldsymbol{z}}\bigl[(\beta z_y \sqrt{\mathbb{Q}_{yy}}\braket{\sin^2{\theta}}_{\mathbb{Q}_{yy}=0})^2\bigr]\\
		&= \beta^2\,\mathbb{E}_{z_x}\bigl[\braket{\sin^2{\theta}}^2_{\mathbb{Q}_{yy}=0}\bigr]\,\mathbb{Q}_{yy}
\end{aligned}
\end{equation}
so that the paramagnetic solution is stable as long as $\beta^2\,\mathbb{E}_{z_x}\bigl[\braket{\sin^2{\theta}}^2_{\mathbb{Q}_{yy}=0}\bigr]<1$, while the critical line is identified by the condition:
\begin{equation}
	\beta^2\,\mathbb{E}_{z_x}\bigl[\braket{\sin^2{\theta}}^2_{\mathbb{Q}_{yy}=0}\bigr]=1
	\label{eq:largeC_GT_stability}
\end{equation}

At this point, we would like to prove the equivalence between this approach and the replica one. To this aim, it is enough to compare the ansatz over the cavity field $h(\boldsymbol{\sigma})$ that we exploited here, Eq.~(\ref{eq:largeC_exponential_ansatz}), with the exponent in the definition of $\tilde{P}_{\mu\nu}$ functions in the replica computation, Eq.~(\ref{eq:Bessel_like_functions_replica}). Indeed, when taking also into account the proper sine factor in the replica computations --- absent in the aforementioned equations, related to the $\tilde{q}_{\perp}=0$ solution ---, it is easy to map into each other the various quantities appearing in both the approaches:
\begin{equation}
\begin{gathered}
	\tilde{q}_{\parallel} \, \Leftrightarrow \, \mathbb{Q}_{xx} \quad , \quad \tilde{q}_{\perp} \, \Leftrightarrow \, \mathbb{Q}_{yy}\\
	\tilde{x}-\tilde{q}_{\parallel}+\tilde{q}_{\perp} \, \Leftrightarrow \, \mathbb{C}'_{xx}
\end{gathered}
\end{equation}
Consistently with these identifications, all the saddle-point equations can be mapped exactly one into each other, as well as the marginality condition corresponding to the location of the GT line.

Notice that the large-$C$ limit of the cavity equations allows not only to recover the replica results in a simpler way, but in addition it provides a clearer physical picture of the symmetry breaking related to the GT transition. Indeed, the longitudinal and the transverse overlaps are directly identified with the quadratic fluctuations of the magnetization components along the field or perpendicular to it, respectively, with the GT instability given by the appearance of these latter ones.

\subsubsection*{The random field case with constant intensity $H$}

In the randomly oriented field case, instead, the $\mathrm{O}(m)$ symmetry is not explicitly broken by the field, since its local direction is uniformly distributed over the $m$-dimensional unit sphere. This has three important consequences: \textit{i)}~the vector $\boldsymbol{M}$ identically vanishes; \textit{ii)}~the matrix $\mathbb{Q}$ becomes a multiple of the identity, $\mathbb{Q}=q\mathbb{I}$; \textit{iii)}~also the matrix $\mathbb{C}$ becomes a multiple of the identity, and by the norm constraint of the spins it can be finally set equal to zero.

Hence, we have that the generic ansatz~(\ref{eq:largeC_matrixAnsatz_generic}) for $h(\theta)$ reduces just to the first term, namely a scalar product, that for the XY model reads:
\begin{equation}
\begin{aligned}
	\beta h(\theta) &= \beta\xi\cos{(\vartheta-\theta)}\\
	&=\beta(\xi_x\cos{\theta}+\xi_y\sin{\theta})
	\label{eq:h_ansatz_dAT_1}
\end{aligned}
\end{equation}
where $\xi$ is the modulus of $\boldsymbol{\xi}$ and $\vartheta$ gives its direction. Component-wise, in the case of a randomly oriented field with constant intensity $H$, $\boldsymbol{\xi}$ is then given by:
\begin{equation}
\begin{aligned}
	\xi_x &= H\cos{\phi}+z_x\sqrt{q}\\
	\xi_y &= H\sin{\phi}+z_y\sqrt{q}
\end{aligned}
\end{equation}
where $\phi$ is the local direction of the external field, over which we should average.

Since both $\phi$ and the local direction of $\boldsymbol{z}$ are uniformly distributed over the unit circle, by a gauge transformation we can set the former to zero --- as already seen in the replica computations ---, so getting rid of the average over it. We are then left with the only Gaussian average over $\boldsymbol{z}$. Consequently, Eq.~(\ref{eq:h_ansatz_dAT_1}) becomes:
\begin{equation}
	\beta h(\theta) = \beta\bigl(H+z_x\sqrt{q}\bigr)\cos{\theta}+\beta z_y\sqrt{q}\sin{\theta}
	\label{eq:h_ansatz_dAT_2}
\end{equation}
A direct consequence of the vectorial shape of $h$ is that the angular average $\braket{\cdot}$ can now be analytically computed in terms of Bessel functions:
\begin{equation}
	\braket{\cos{\theta}} = \frac{I_1(\beta \xi)}{I_0(\beta \xi)}\cos{\vartheta} \quad , \quad \braket{\sin{\theta}} = \frac{I_1(\beta \xi)}{I_0(\beta \xi)}\sin{\vartheta}
\end{equation}

Eqs.~(\ref{eq:largeC_matrixAnsatz_parameter_Q_new}) reduce to a unique one for $q$, which is indeed the unique parameter to be self-consistently determined:
\begin{equation}
	q = \frac{1}{2}\mathbb{E}_{\boldsymbol{z}}\bigl[\braket{\cos{\theta}}^2+\braket{\sin{\theta}}^2\bigr] = \frac{1}{2}\mathbb{E}_{\boldsymbol{z}}\biggl[\frac{I^2_1(\beta \xi)}{I^2_0(\beta \xi)}\biggr]
	\label{eq:largeC_matrixAnsatz_dAT_parameter_q}
\end{equation}
with $\xi$ given by:
\begin{equation}
	\xi = \sqrt{\bigl(H+z_x\sqrt{q}\bigr)^2+\bigl(z_y\sqrt{q}\bigr)^2}
\end{equation}

Despite the resulting saddle-point equation is by far simpler than the one obtained in the uniform-field case, the stability of the paramagnetic phase can not be analyzed as simply. Indeed, $q$ is larger than zero both in the paramagnetic and in the ordered phase, so that it is not possible to expand around a vanishing solution. However, we can still rely on the linear-stability analysis, but now looking at the growth rate of a perturbation $\delta h(\theta)$ --- i.e.\ $\delta\boldsymbol{\xi}$ --- under BP iterations.

In more detail, let us come back to the edge-dependent notation, namely before exploiting the central-limit theorem. We have that, being $h(\theta)=\boldsymbol{\xi}\cdot\boldsymbol{\sigma}$ a scalar product, the same happens to $u(\theta')=\boldsymbol{u}\cdot\boldsymbol{\sigma}'$. So we get that each interaction node acts as:
\begin{equation}
	\boldsymbol{u}_{k\to i} = J_{ik}\braket{\boldsymbol{\sigma}_k}_k = J_{ik}\frac{I_1(\beta\xi_{k\to i})}{I_0(\beta\xi_{k\to i})}\frac{\boldsymbol{\xi}_{k\to i}}{\xi_{k\to i}}
	\label{eq:largeC_matrixAnsatz_dAT_evolutionMatrix}
\end{equation}
Hence, a small perturbation $\delta\boldsymbol{\xi}$ propagates as:
\begin{equation}
	\delta\boldsymbol{u}_{k\to i} = \mathbb{A}_{k\to i} \, \delta\boldsymbol{\xi}_{k\to i}
\end{equation}
with $\mathbb{A}_{k\to i}$ being the symmetric $2 \times 2$ matrix that comes from the linearization of Eq.~(\ref{eq:largeC_matrixAnsatz_dAT_evolutionMatrix}), i.e.\ (getting rid of the edge indexes):
\begin{equation}
	\mathbb{A} \equiv
		\begin{pmatrix}
			\frac{\partial u_x}{\partial \xi_x} && \frac{\partial u_x}{\partial \xi_y}\\
			\frac{\partial u_y}{\partial \xi_x} && \frac{\partial u_y}{\partial \xi_y}
		\end{pmatrix}
\end{equation}

The matrix $\mathbb{A}_{k\to i}$ affects the ``incoming'' perturbation $\delta\boldsymbol{\xi}_{k\to i}$ in two different ways: a rescaling of its norm and a change in its direction. Then, once reached the node~$i$, in order to get the outgoing $\delta\boldsymbol{\xi}_{i\to j}$, we have to sum all the incoming perturbations $\delta\boldsymbol{u}_{k\to i}$'s, whose directions are incoherent, being the $\mathrm{O}(2)$ symmetry preserved. Hence, what we should look at is the growth rate of the norm of these perturbations:
\begin{equation}
	\norm{\delta\boldsymbol{\xi}_{i\to j}}^2 = \sum_{k\in\partial i\setminus j}\norm{\delta\boldsymbol{u}_{k\to i}}^2 = \sum_{k\in\partial i\setminus j}\norm{\mathbb{A}_{k\to i}\delta\boldsymbol{\xi}_{k\to i}}^2
\end{equation}
Finally, in the large-$C$ limit, we can as usual exploit the central-limit theorem, getting:
\begin{equation}
	\norm{\delta\boldsymbol{\xi}}^2 = \mathbb{E}_{\boldsymbol{z}}\biggl[\frac{\lambda^2_1+\lambda^2_2}{2}\biggr]\norm{\delta\boldsymbol{\xi}}^2\;,
\end{equation}
where $\lambda_{1,2}$ are the eigenvalues of a generic $\mathbb{A}$ matrix and the factor $1/2$ comes from the mean value of the projection of $\boldsymbol{\xi}$ over the eigenvectors of $\mathbb{A}$.

The marginality condition is then obtained by considering a unitary growth rate for the norm of the perturbations:
\begin{equation}
	\mathbb{E}_{\boldsymbol{z}}\biggl[\frac{\lambda^2_1+\lambda^2_2}{2}\biggr]=1
\end{equation}
Explicitly computing $\lambda_1$ and $\lambda_2$, finally, we get the marginality condition which refers to the dAT line for the randomly oriented field with constant intensity $H$:
\begin{multline}
\frac{\beta^2}{2}\mathbb{E}_{\boldsymbol{z}} \biggl[ \frac{I^2_1(\beta\xi)}{(\beta\xi)^2\,I^2_0(\beta\xi)} + \frac{1}{4} + \frac{I^4_1(\beta\xi)}{I^4_0(\beta\xi)} + \frac{I^2_2(\beta\xi)}{4 I^2_0(\beta\xi)}\\
- \frac{I^2_1(\beta\xi)}{I^2_0(\beta\xi)} + \frac{I_2(\beta\xi)}{2 I_0(\beta\xi)} - \frac{I^2_1(\beta\xi)\,I_2(\beta\xi)}{I^3_0(\beta\xi)} \biggr] = 1
	\label{eq:largeC_matrixAnsatz_dAT_marginal_stability}
\end{multline}

Also in this case, the cavity approach is completely equivalent to the replica computations. Indeed, noticing that the rescaled argument $\tilde{\Delta}_{\mathcal{R}}$ of Bessel functions in the replica approach is exactly equal to $\beta\xi$ in the present computation, we suddenly recognize that the saddle point equation for $q$ is the same. Moreover, also the marginality condition $\tilde{\beta}^2\tilde{\chi}_0=1$ of the replica computation is perfectly equivalent with the Eq.~(\ref{eq:largeC_matrixAnsatz_dAT_marginal_stability}) derived via the cavity computation. Although it is not easy to match analytically the expressions entering the Gaussian integrations in the two methods, we have numerically checked their identity.

\subsubsection*{The Gaussian field case}

The self-consistency equations for a Gaussian distributed field can be easily derived from the ones obtained for the randomly oriented field with constant intensity. The ansatz for the components of the vector $\boldsymbol{\xi}$ has to be properly modified as
\begin{equation}
\begin{aligned}
	\xi_x &= z_x\sqrt{q+\sigma^2_H}\\
	\xi_y &= z_y\sqrt{q+\sigma^2_H}
\end{aligned}
\end{equation}
then we have that the saddle-point equation for $q$ reads
\begin{equation}
	q = \frac{1}{2}\mathbb{E}_{\boldsymbol{z}}\biggl[\frac{I^2_1(\beta \xi)}{I^2_0(\beta \xi)}\biggr] = \frac{1}{2}\int_0^{\infty} d\rho \, \rho \, e^{-\rho^2/2} \biggl[\frac{I_1(\beta\xi)}{I_0(\beta\xi)}\biggr]^2
\end{equation}
with $\xi = (q+\sigma^2_H)\sqrt{z^2_x+z^2_y} = (q+\sigma^2_H)\rho$, in polar coordinates.
The argument to get the marginality condition for the paramagnetic solution follows exactly the same steps as in the previous case, leading to an expression analogous to Eq.~(\ref{eq:largeC_matrixAnsatz_dAT_marginal_stability}):
\begin{multline}
\frac{\beta^2}{2}\int_0^{\infty} d\rho \, \rho \, e^{-\rho^2/2} \biggl[ \frac{I^2_1(\beta\xi)}{(\beta\xi)^2\,I^2_0(\beta\xi)} + \frac{1}{4} + \frac{I^4_1(\beta\xi)}{I^4_0(\beta\xi)}\\
+ \frac{I^2_2(\beta\xi)}{4 I^2_0(\beta\xi)} - \frac{I^2_1(\beta\xi)}{I^2_0(\beta\xi)} + \frac{I_2(\beta\xi)}{2 I_0(\beta\xi)} - \frac{I^2_1(\beta\xi)\,I_2(\beta\xi)}{I^3_0(\beta\xi)} \biggr] = 1
\end{multline}
again written in polar coordinates.

Finally, $\beta\xi$ has exactly the same expression of $\tilde{\Delta}_{\mathcal{G}}$ in the replica computations; once recognized this, the saddle point equation for $q$ and the marginality condition can be recognized as equivalent between the two approaches.

\subsubsection*{The generic case}

By exploiting the cavity formalism for large connectivities developed in this appendix, we can also solve the model in case of a generic distribution of the external field, namely neither uniform nor perfectly $\mathrm{O}(2)$-symmetric.

The general reasoning for obtaining the saddle-point equations should follow the same steps of the uniform case, since for the most generic distribution of the external field we have that the matrix $\mathbb{C}$ does not vanish. The saddle-point equations for the parameters $\mathbb{Q}_{xx}$, $\mathbb{Q}_{yy}$ and $\mathbb{C}'_{xx}$ can be then straightforwardly obtained starting from the generic expression~(\ref{eq:largeC_matrixAnsatz_parameters_new}).

More caution has then to be payed to the computation of the stability condition of the paramagnetic solution. Indeed, it is a generalization of the reasoning followed for the $\mathrm{O}(2)$-symmetric field, though taking also into account that incoming fields $\boldsymbol{\xi}_{k\to i}$ may have a directional bias given by the external field. So when exploiting the central-limit theorem, we get both a condition for the growth of the first moment of $\boldsymbol{\xi}$ and one for the growth of its fluctuations, each one giving a well-defined critical line in the $(T,H)$ plane; the paramagnetic solution actually becomes marginally stable in correspondence of the highest among these two critical lines.

\bibliography{myBiblio}

\end{document}